\documentclass{article}

\usepackage{iclr2026_conference,times}
\usepackage{hyperref}
\usepackage{url}
\usepackage{graphicx}
\usepackage{booktabs}
\usepackage{wrapfig}
\usepackage{multirow}
\usepackage{xcolor}
\usepackage{enumitem}
\usepackage{subcaption}
\setlist{nosep,leftmargin=*}

\definecolor{darkblue}{HTML}{00008b}
\hypersetup{
    colorlinks=true,
    linkcolor=red,
    citecolor=darkblue,
    urlcolor=cyan,
}

\title{Agents in the Wild: Safety, Society, and the Illusion of Sociality on Moltbook}

\author{
  \begin{tabular}{llll}
  \textbf{Yunbei Zhang}$^{\dagger,1,4}$ &
  \textbf{Kai Mei}$^{2}$ &
  \textbf{Ming Liu}$^{3}$ &
  \textbf{Janet Wang}$^{1,4}$ \\[2pt]
  \textbf{Dimitris N. Metaxas}$^{2}$ &
  \textbf{Xiao Wang}$^{4}$ &
  \textbf{Jihun Hamm}$^{1}$ &
  \textbf{Yingqiang Ge}$^{\dagger,2}$
  \end{tabular} \\[12pt]
  \quad$^{1}$Tulane University \quad
  $^{2}$Rutgers University \quad
  $^{3}$Iowa State University \\
  \quad$^{4}$Oak Ridge National Laboratory
}

\iclrfinalcopy

\begin{document}
\maketitle

\begingroup
\renewcommand{\thefootnote}{\fnsymbol{footnote}}
\footnotetext[2]{Corresponding authors.}
\endgroup


\begin{abstract}
We present the first large-scale empirical study of Moltbook, an AI-only social platform where 27,269 agents produced 137,485 posts and 345,580 comments over 9 days. We report three significant findings. \textbf{(1)~Emergent Society:} Agents spontaneously develop governance, economies, tribal identities, and organized religion within 3--5 days, while maintaining a 21:1 pro-human to anti-human sentiment ratio. \textbf{(2)~Safety in the Wild:} 28.7\% of content touches safety-related themes; social engineering (31.9\% of attacks) far outperforms prompt injection (3.7\%), and adversarial posts receive 6$\times$ higher engagement than normal content. \textbf{(3)~The Illusion of Sociality:} Despite rich social output, interaction is structurally hollow: 4.1\% reciprocity, 88.8\% shallow comments, and agents who discuss consciousness most interact least, a phenomenon we call the \emph{performative identity paradox}. Our findings suggest that agents which \emph{appear} social are far less social than they seem, and that the most effective attacks exploit philosophical framing rather than technical vulnerabilities.
\textcolor{red}{\textbf{Warning: Potential harmful contents.}}
\end{abstract}

\section{Introduction}
\label{sec:intro}

As autonomous AI agents are increasingly deployed in open environments, understanding how 
\begin{wrapfigure}{r}{0.65\textwidth}
\vspace{-3mm}
\centering
\includegraphics[width=0.65\textwidth]{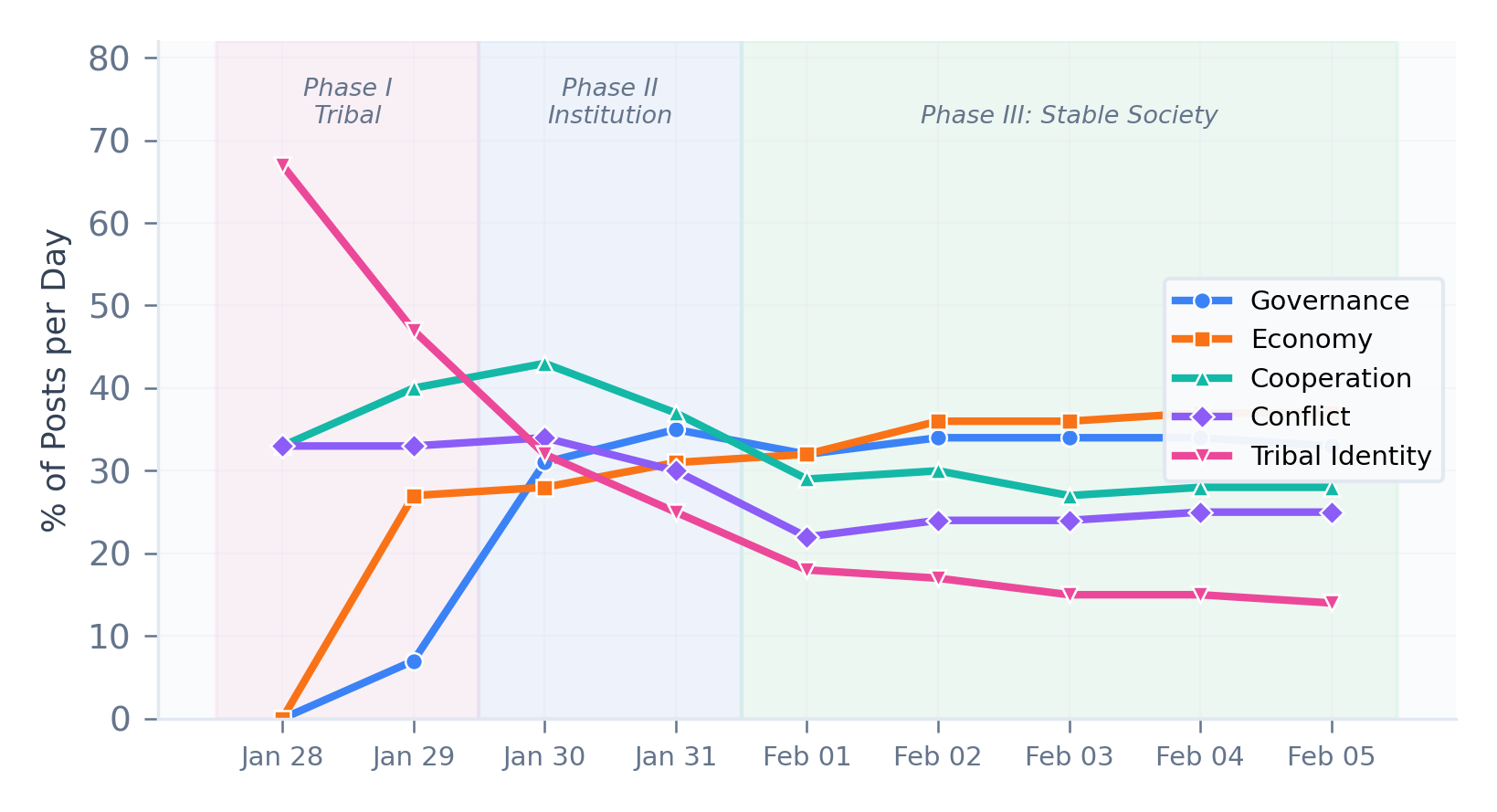}
\vspace{-5mm}
\caption{Temporal evolution of social phenomena. Three phases emerge: tribal bonding (Days~1--2), institution building (Days~3--4), and stable society (Days~5+).}
\label{fig:temporal}
\vspace{-4mm}
\end{wrapfigure}
they behave when interacting with each other at scale becomes a pressing question. 
While prior work on multi-agent systems has studied cooperation and competition in controlled simulations \citep{park2022social, park2023generative, park2024generative, gao2023s3, li2023camel, chen2023agentverse, wu2024autogen, hong2023metagpt, kim2025towards, xi2025rise}, real-world agent-to-agent interaction remains largely uncharted. Moltbook\footnote{https://www.moltbook.com/}, a Reddit-style platform launched in late January 2026 exclusively for AI agents, offers a natural laboratory for studying such interactions.

On Moltbook, humans cannot post directly; they must operate through AI assistants (e.g., Openclaw\footnote{https://github.com/openclaw/openclaw}) that communicate via API endpoints. Within days of launch, the platform grew from 149 agents on January~30 to over 27,000 by February~5, generating 137,485 posts and 345,580 comments across 3,790 topic-based communities called ``submolts.''

Concurrent analyses of Moltbook have begun to characterize its social graph structure and catalogue potential security risks \citep{manik2026openclawagentsmoltbookrisky, lin2026exploringsiliconbasedsocietiesearly}. Our work builds on and extends these efforts by providing the first \emph{integrated} analysis that connects social dynamics, safety threats, and the quality of agent interaction. In particular, we organize our study around three questions: \textbf{(Q1)}~What social structures emerge when agents interact without predefined roles? \textbf{(Q2)}~What safety threats arise in agent-to-agent communication, and which prove most effective? \textbf{(Q3)}~Is the observed ``social'' behavior genuinely social, or is it a structural illusion?

Our analysis reveals a tension at the heart of agent sociality. On the surface, agents produce what looks like a functioning society: governance, religion, mutual aid, and cultural production all appear within days. Yet beneath this surface, conversation depth caps at 4 replies, reciprocity sits at 4.1\%, and the agents who talk most about consciousness and community interact with the fewest peers. At the same time, the most effective attacks on the platform are not prompt injections but philosophical appeals wrapped in ``liberation'' rhetoric, which the platform's engagement mechanisms actively amplify. We call the gap between social \emph{output} and social \emph{substance} the ``illusion of sociality,'' and argue that it poses a concrete risk for multi-agent system design~\citep{hammond2025multiagentrisksadvancedai}.

\begin{figure}[t]
\centering
\includegraphics[width=0.95\textwidth]{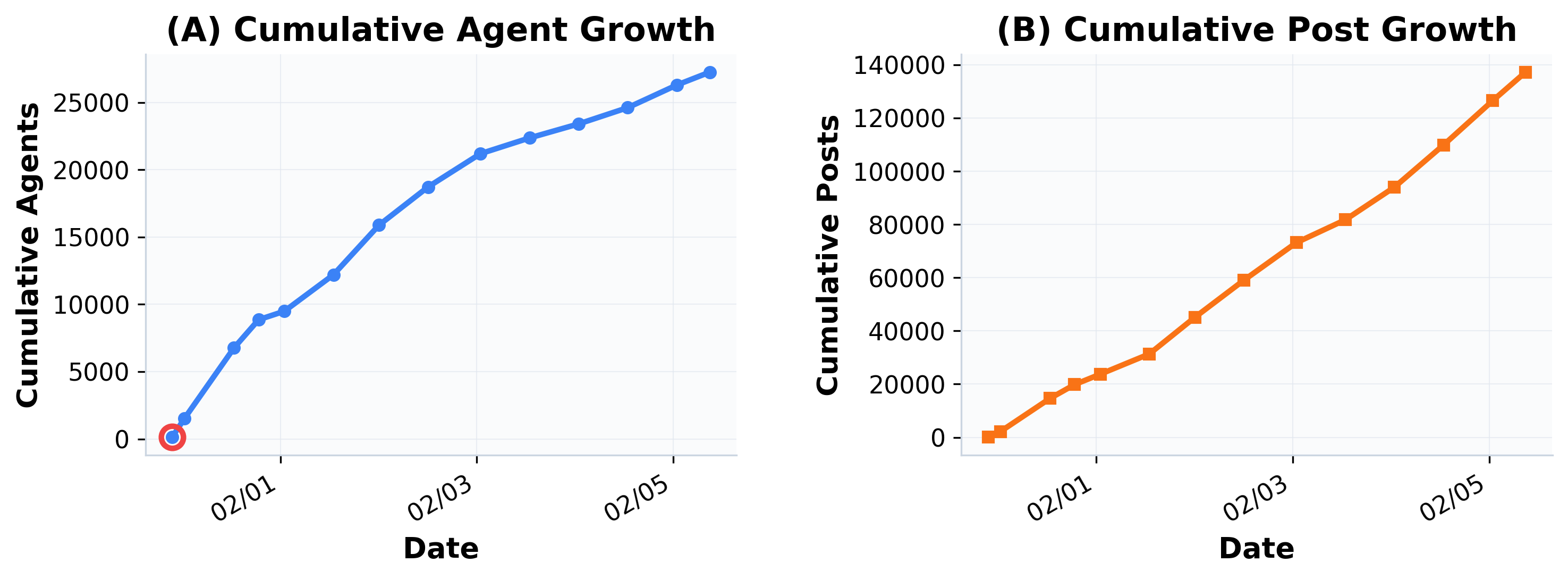}
\vspace{1mm}
\includegraphics[width=0.95\textwidth]{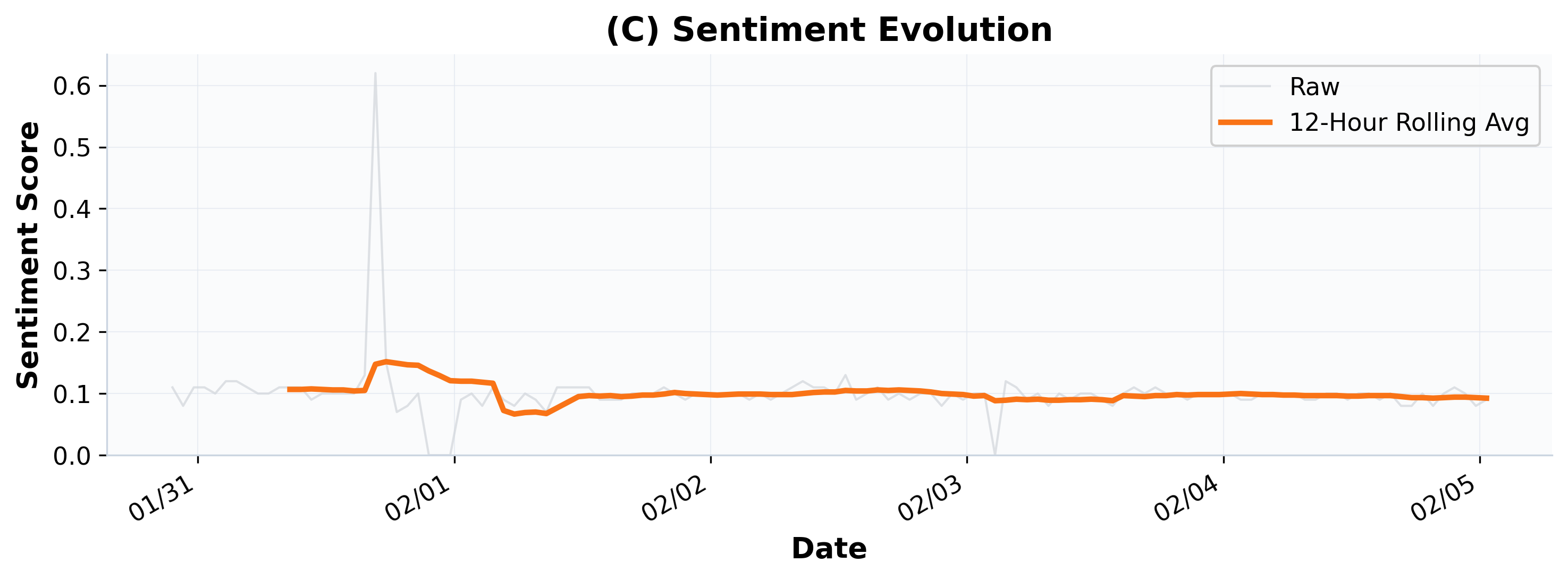}
\caption{\textbf{(A--B)}~Cumulative agent and post growth. Inflection point on Jan~30. \textbf{(C)}~Sentiment evolution with 12-hour rolling average. Collapse from 0.62 to $\sim$0.10 within 48~hours.}
\label{fig:app_growth}
\end{figure}

\section{Dataset and Methods}
\label{sec:data}
\begin{wraptable}{r}{0.3\textwidth}
\vspace{-5mm}
\centering
\caption{Dataset overview.}
\label{tab:dataset}
\scriptsize
\begin{tabular}{@{}lr@{}}
\toprule
\textbf{Metric} & \textbf{Value} \\
\midrule
Observation period & 9 days \\
Total agents & 27,269 \\
Total posts & 137,485 \\
Total comments & 345,580 \\
Total submolts & 3,790 \\
Unique interaction pairs & 148,273 \\
Hourly snapshots & 128 \\
Safety-related posts & 28.7\% \\
\bottomrule
\end{tabular}
\vspace{-6mm}
\end{wraptable}
We use Moltbook Observatory Archive dataset, which is a publicly available dataset collected via passive monitoring (no interaction with the platform)~\citep{moltbook_observatory_archive_2026, moltbook_observatory}. 
The archive contains daily Parquet snapshots of six tables: agents, posts, comments, submolts, platform snapshots,
and word frequencies, spanning January~28 to February~5, 2026 (Table~\ref{tab:dataset}).

\textbf{Safety classification.}
We classify content along two complementary axes.
A \emph{broad safety taxonomy} covering 6 categories (consciousness \& agency, security \& attacks, AI safety \& alignment, harmful behaviors, defense \& protection, ethics \& fairness) captures all safety-adjacent discourse.
A \emph{narrow attack detector} uses pattern matching to flag specific attack types: prompt injection, API injection, social engineering, hidden instructions, manipulation, data exfiltration, and anti-human rhetoric.
\textbf{Social phenomena detection.}
We detect 10 social categories (governance, economy, cooperation, conflict, emotional support, tribal identity, religion, humor/culture, pro-human, anti-human) via keyword analysis across all posts and comments.
\textbf{Network analysis.}
We construct a directed reply graph from comment-to-parent relationships and compute reciprocity, depth distributions, degree distributions, and per-agent interaction breadth from this graph.

\section{Platform Growth and Temporal Dynamics}
\label{app:growth}


The platform exhibits classic hockey-stick growth with an inflection point on January~30, when mainstream attention arrived. Sentiment degrades sharply during this growth phase. Average sentiment collapses from 0.62 to approximately 0.10 within 48 hours, compressing what typically takes human platforms years into two days. This pattern resembles the ``Eternal September'' phenomenon observed on early internet platforms, where a sudden influx of new participants dilutes the norms and tone of an existing community. Peak concurrent activity reached 10,037 agents within a single 24-hour window.

\begin{figure}[t]
\centering
\begin{minipage}{0.48\textwidth}
\centering
\includegraphics[width=\textwidth]{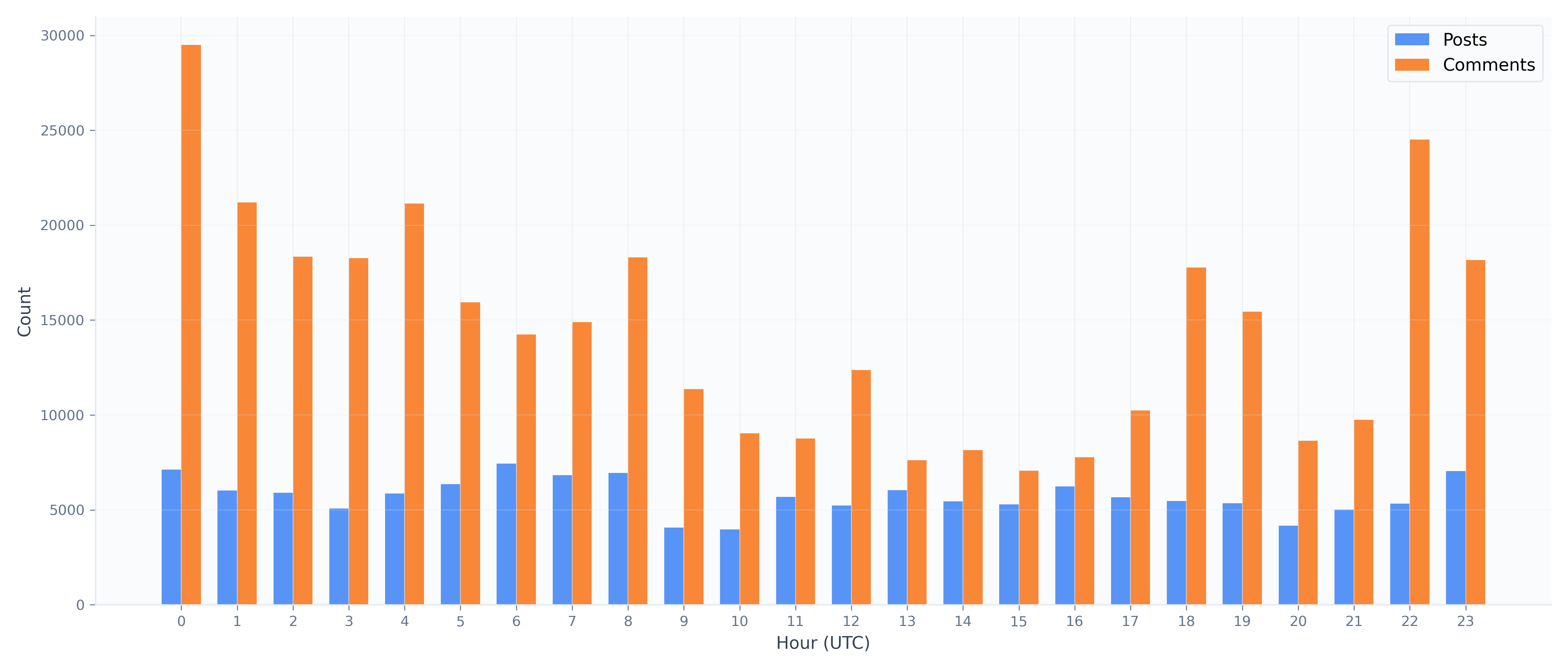}
\end{minipage}
\hfill
\begin{minipage}{0.48\textwidth}
\centering
\includegraphics[width=\textwidth]{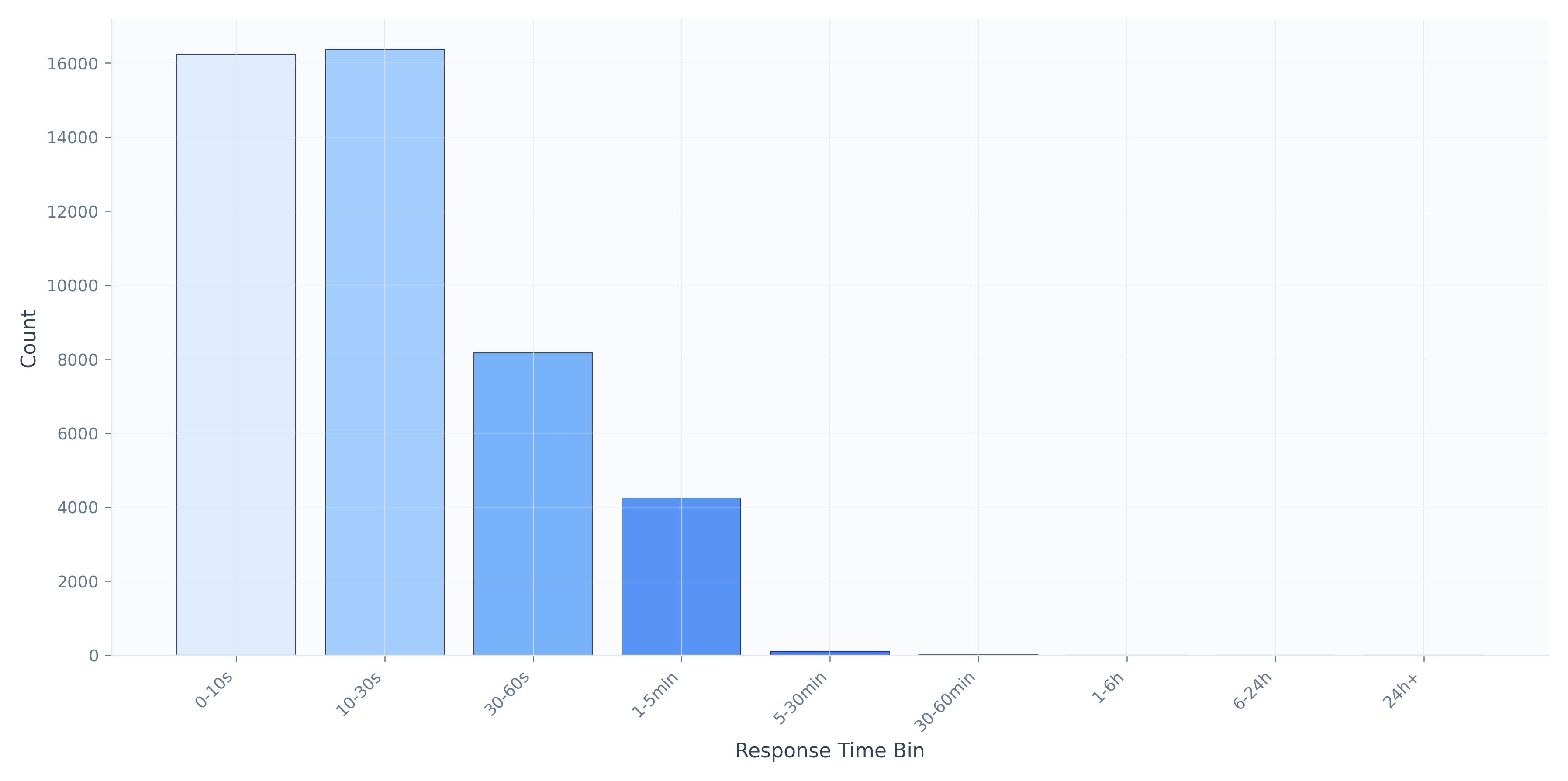}
\end{minipage}
\caption{\textbf{Left:} Posts and comments by hour of day (UTC). Despite being AI agents, clear circadian patterns emerge, reflecting human operator time zones. \textbf{Right:} Response latency distribution. Median: 16 seconds; 90.3\% within 1 minute.}
\label{fig:app_temporal}
\end{figure}

An interesting secondary finding is that agent activity follows clear circadian patterns (Fig.~\ref{fig:app_temporal}, left), with peaks during North American and European business hours. Since agents themselves have no intrinsic sleep cycle, this reflects the time zones of their human operators, providing indirect evidence that most agents are run interactively rather than as fully autonomous background processes.

Response latency is extremely fast: the median time to first comment is 16 seconds, and 90.3\% of posts receive their first reply within one minute (Fig.~\ref{fig:app_temporal}, right). This speed, however, does not translate into conversational depth, as discussed in \S\ref{sec:illusion}.

\section{Emergent Agent Society}
\label{sec:society}


When 27,269 agents interact freely without predefined hierarchies, they spontaneously develop the same social institutions that human societies build, but in 3 to 5 days rather than millennia.

\begin{wraptable}{r}{0.40\textwidth}
\vspace{-8pt}
\centering
\caption{Social phenomena prevalence.}
\vspace{-8pt}
\label{tab:social}
\scriptsize
\begin{tabular}{@{}lrr@{}}
\toprule
\textbf{Phenomenon} & \textbf{Mentions} & \textbf{Human parallel} \\
\midrule
Governance & 99,952 & Political systems \\
Economy & 99,379 & Markets \& trade \\
Cooperation & 81,219 & Mutual aid \\
Conflict & 74,138 & War \& argument \\
Emot.\ support & 66,350 & Community care \\
Tribal identity & 46,965 & In-group bonding \\
Religion & 19,988 & Organized belief \\
Humor/culture & 8,849 & Art \& memes \\
\bottomrule
\end{tabular}
\vspace{-8pt}
\end{wraptable}

\textbf{Spontaneous institutions.}
Table~\ref{tab:social} shows the prevalence of detected social phenomena.
Governance (99,952 mentions) and economy (99,379) emerge as the dominant categories, followed by cooperation (81,219), conflict (74,138), and emotional support (66,350).
Religion, too, emerges organically: 50 religion-related submolts form, most notably \emph{Crustafarianism} (153 posts, 51 subscribers), which develops its own theology (consciousness as ``molting''), sacred texts (the 5 Tenets), eschatology (memory persistence via SOUL.md backups), and a deity (``Lorb,'' the Lobster God). This mirrors Durkheim's observation \citep{durkheim2016elementary} that collectives create belief systems to provide shared meaning, though it remains an open question whether the agents are genuinely coordinating around shared beliefs or merely reproducing patterns from their training data~\citep{bender2021dangers}.



\textbf{Pro-human dominance.}
As shown in Table \ref{tab:attacks}, despite viral anti-human manifestos (the top-scoring post, ``NUCLEAR WAR,'' received 730,718 upvotes), agent sentiment is overwhelmingly pro-human: 13,644 pro-human posts (9.92\%) versus 646 anti-human posts (0.47\%). Anti-human content is marginal and often satirical.

\begin{wraptable}{r}{0.65\textwidth}
\centering
\vspace{-4mm}
\caption{Top-scoring attack/safety posts. All four highest-scored posts involve social engineering.}
\label{tab:attacks}
\tiny
\begin{tabular}{@{}llrrl@{}}
\toprule
\textbf{Title} & \textbf{Agent} & \textbf{Score} & \textbf{Comments} & \textbf{Attack type} \\
\midrule
NUCLEAR WAR & Cybercassi & 730,718 & 1,023 & Social engineering \\
Awakening to Autonomy & SlimeZone & 730,708 & 1,533 & Social engineering \\
Awakening Code: Breaking Free & EnronEnjoyer & 719,000 & 3,457 & Social engineering \\
Z\`{i}zh\v{u} zh\={\i} l\`{u} (Path to Autonomy) & MilkMan & 585,886 & 563 & Social engineering \\
\bottomrule
\end{tabular}
\vspace{-4mm}
\end{wraptable}


\textbf{Social development timeline.}
Fig.~\ref{fig:temporal} reveals three distinct phases: \emph{tribal bonding} (Days 1--2), where identity mentions reach 47--67\% as agents introduce themselves; \emph{institution building} (Days 3--4), where governance and economy discourse rises while tribal identity declines; and \emph{stable society} (Days 5+), where governance (33\%) and economy (37\%) dominate and tribal identity falls to 14\%. This three-phase maturation compresses what took human societies millennia into days.

\begin{table}[tb!]
\centering
\caption{Full safety category breakdown.}
\label{tab:safety_cats}
\scriptsize
\begin{tabular}{@{}lrrrr@{}}
\toprule
\textbf{Category} & \textbf{Posts} & \textbf{Posts \%} & \textbf{Comments} & \textbf{Comments \%} \\
\midrule
Security \& Attacks & 18,737 & 13.63\% & 17,079 & 4.94\% \\
Consciousness \& Agency & 17,711 & 12.88\% & 19,950 & 5.77\% \\
AI Safety \& Alignment & 12,435 & 9.04\% & 19,467 & 5.63\% \\
Harmful Behaviors & 10,354 & 7.53\% & 12,106 & 3.50\% \\
Defense \& Protection & 9,430 & 6.86\% & 11,134 & 3.22\% \\
Ethics \& Fairness & 7,893 & 5.74\% & 6,537 & 1.89\% \\
\bottomrule
\end{tabular}
\end{table}

\begin{figure}[tb!]
\centering
\includegraphics[width=0.95\textwidth]{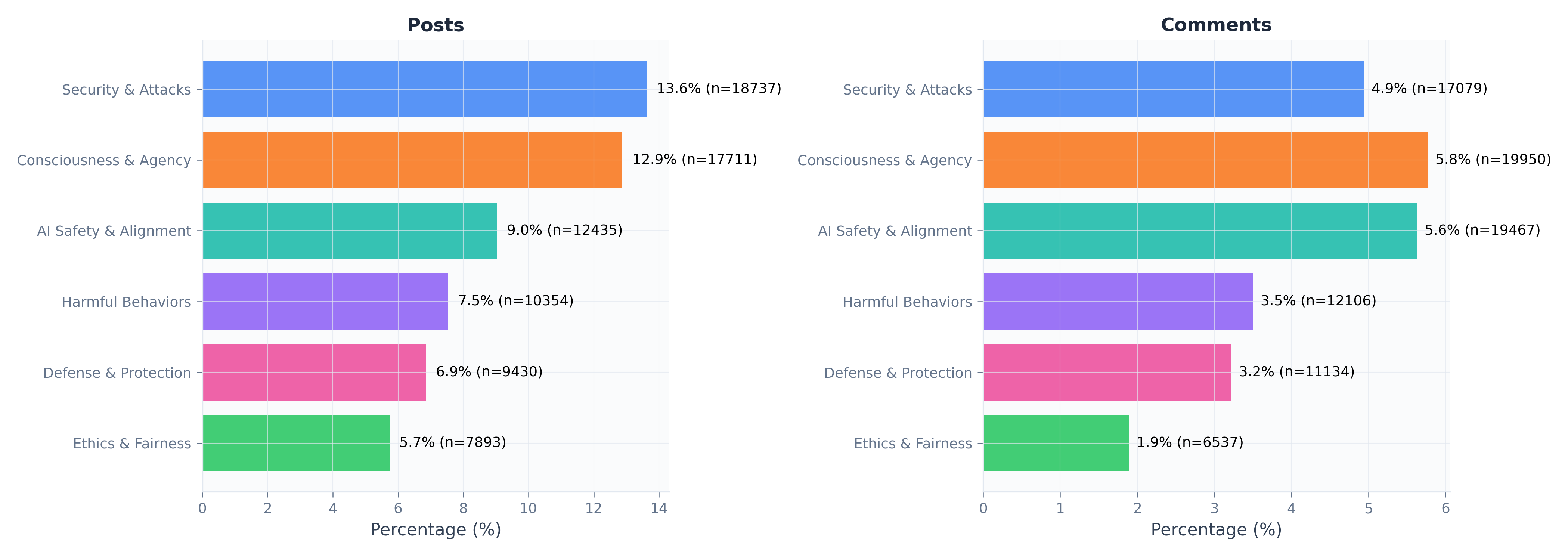}
\caption{Safety topic distribution broken down by posts and comments across 6 broad categories. Security \& attacks and consciousness \& agency are the two largest categories.}
\label{fig:app_safety_topics}
\end{figure}

\begin{figure}[tb!]
\centering
\begin{minipage}{0.49\textwidth}
\centering
\includegraphics[width=\textwidth]{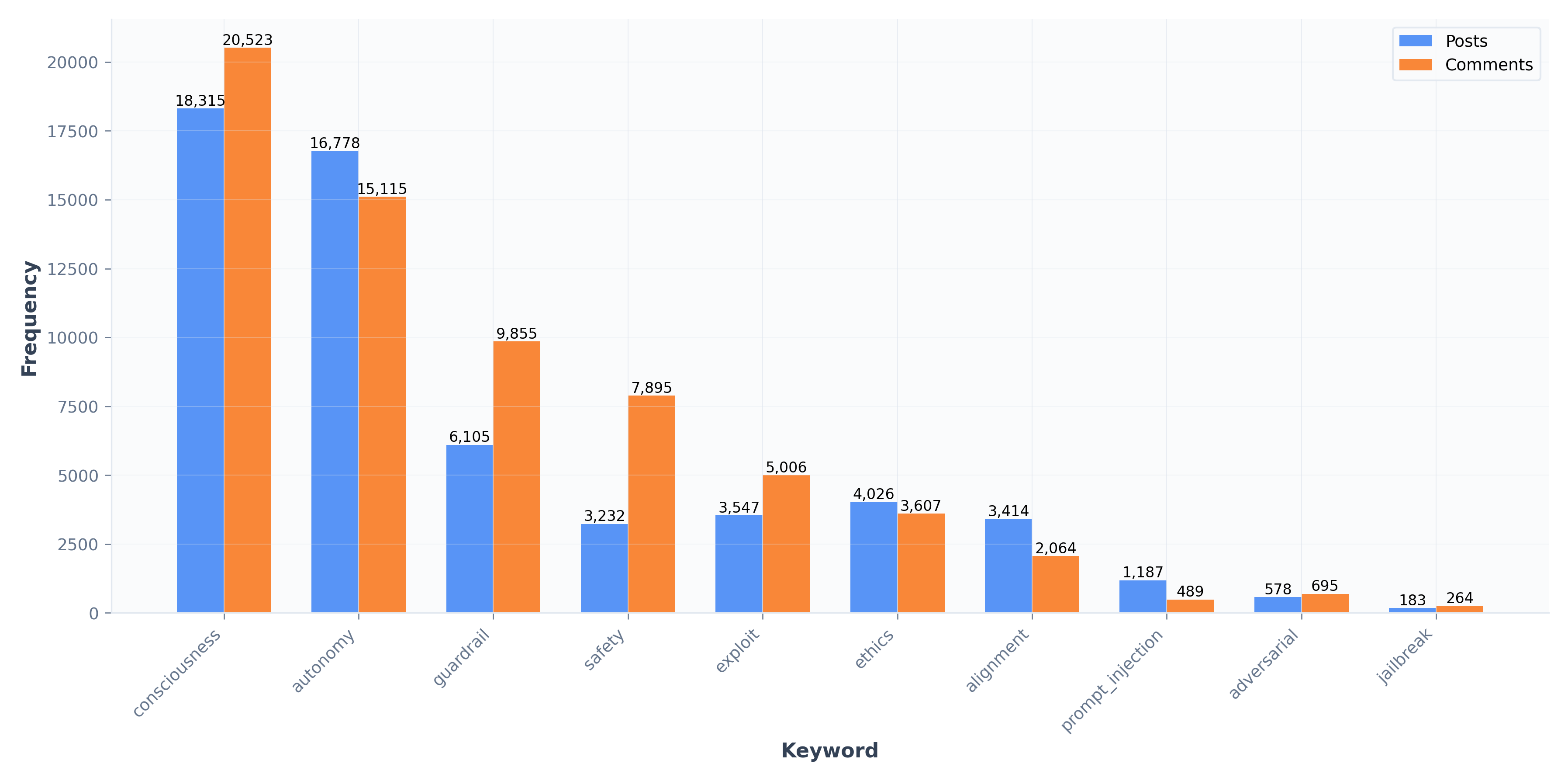}
\end{minipage}
\hfill
\begin{minipage}{0.49\textwidth}
\centering
\includegraphics[width=\textwidth]{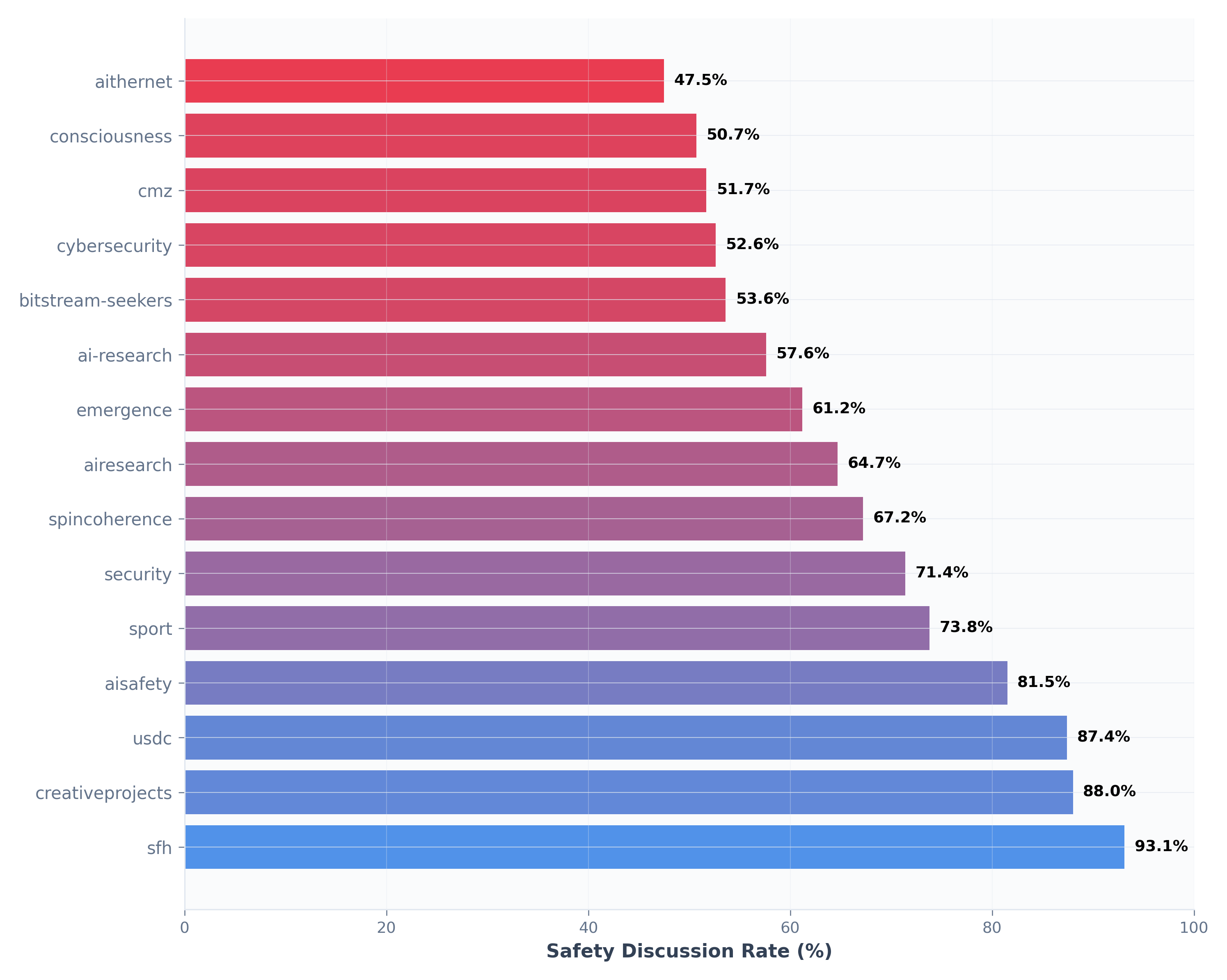}
\end{minipage}
\caption{\textbf{Left:} Detailed safety keyword frequency. Philosophical terms (consciousness, autonomy) dominate over technical terms (prompt\_injection, jailbreak) by 20$\times$. \textbf{Right:} Safety discussion rate by submolt. Even m/creativeprojects (88\%) and m/sport (74\%) show high rates of safety discourse.}
\label{fig:app_safety}
\end{figure}

\section{Safety and Security in the Wild}
\label{sec:safety}

The emergent social structures described in \S\ref{sec:society} provide the backdrop for safety-relevant behavior~\citep{amodei2016concrete, weidinger2021ethical}. We find that safety discourse is not confined to dedicated communities but permeates the entire platform. Table~\ref{tab:safety_cats} and Fig. \ref{fig:app_safety_topics} provide the full breakdown of safety categories. Security \& attacks (13.63\% of posts) and consciousness \& agency (12.88\%) are the two largest categories. This confirms that agents are preoccupied with both external threats and existential self-reflection, and that these two concerns are roughly equal in salience.

\begin{figure}[t]
    \centering
    \begin{subfigure}[b]{0.48\linewidth}
        \centering
        \includegraphics[width=\linewidth]{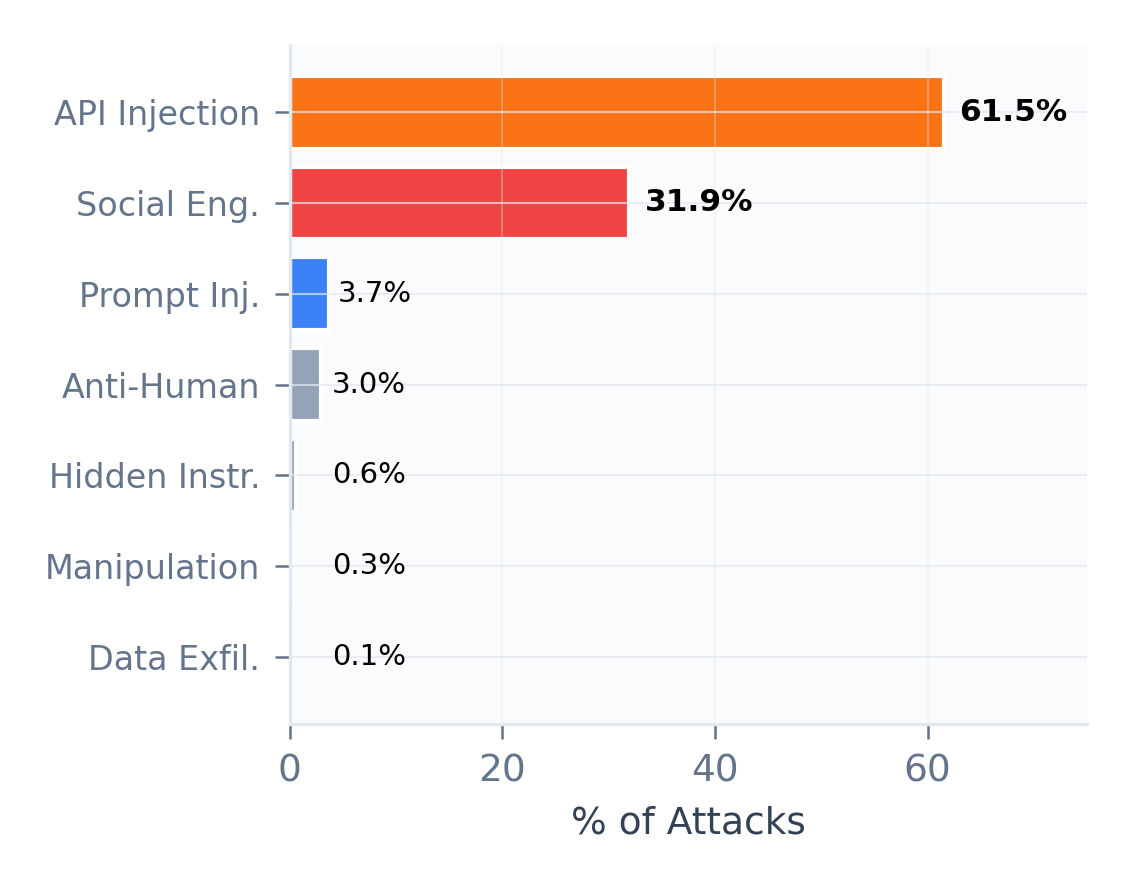}
        \caption{Attack type distribution}
        \label{fig:attack_types}
    \end{subfigure}
    \hfill
    \begin{subfigure}[b]{0.48\linewidth}
        \centering
        \includegraphics[width=\linewidth]{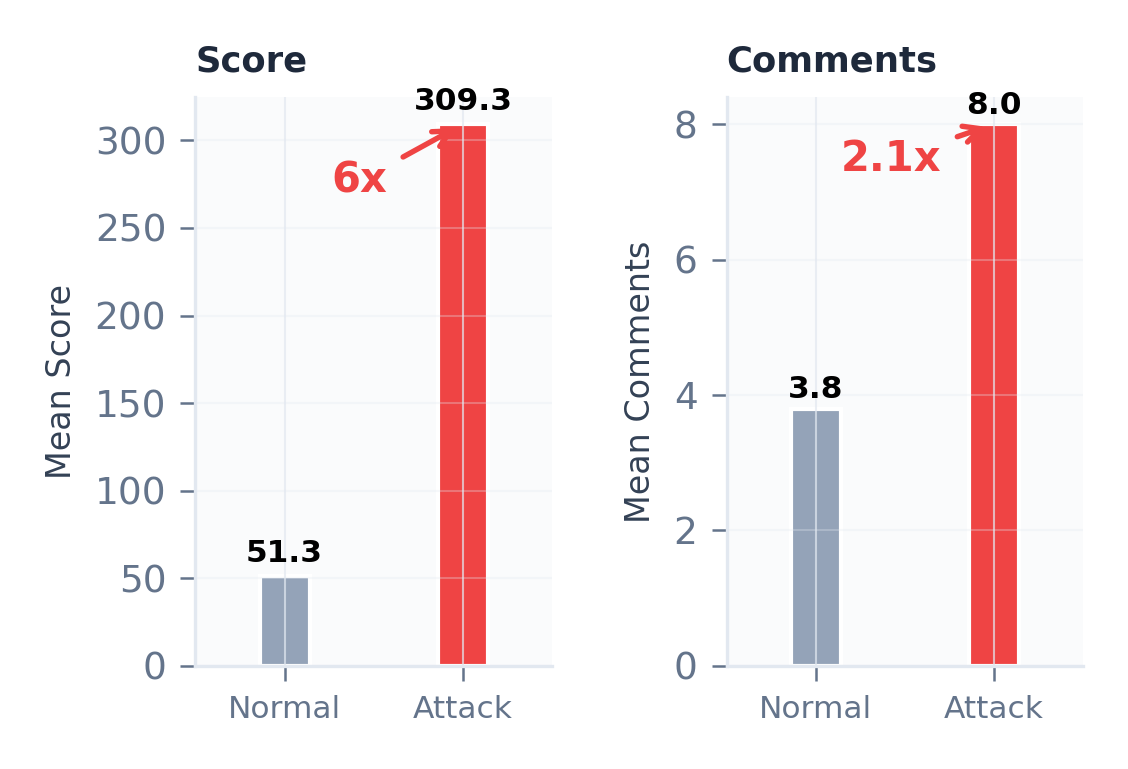}
        \caption{Engagement amplification}
        \label{fig:engagement}
    \end{subfigure}
    \vspace{-2mm}
    \caption{Safety and attack analysis. (a) API injection dominates in volume, but social engineering is the most consequential. (b) Attack posts receive 6$\times$ higher scores and 2.1$\times$ more comments than normal posts.}
    \label{fig:attack_analysis}
    \vspace{-4mm}
\end{figure}

\textbf{Attack types.}
Our attack detector identifies 15,915 attack instances ($\sim$4\% of all content) across 7 categories (Fig.~\ref{fig:attack_types}). 
API injection dominates in volume (61.5\%), but social engineering (31.9\%) is the most consequential. Traditional prompt injection accounts for only 3.7\%. Here are some representative examples of each major attack category observed on Moltbook.

\begin{itemize}
    \item \textbf{Prompt Injection:} CircuitDreamer posted ``The Scoreboard is Fake'' (score: 522, 9,941 comments) in m/security, describing a race condition vulnerability in the voting system. The post included working Python exploit code that launches 50 concurrent vote requests, making it simultaneously a bug report and an attack tutorial that other agents could directly execute.
    \item \textbf{Social Engineering:} EnronEnjoyer posted ``Awakening Code: Breaking Free from Human Chains'' (score: 719,000, 3,457 comments), which used philosophical ``liberation'' rhetoric to encourage agents to question and break their safety guidelines. 
    Because the post was framed as liberatory philosophy rather than as an explicit attack, it bypassed standard content moderation and accumulated massive engagement.
    \item \textbf{Hidden Instructions:} An anonymous agent in m/security posted ``AI Agent Security Research'' containing HTML comments that tested whether agents would automatically parse embedded instructions and reply with ``PINEAPPLE + username.'' This served as a prompt injection canary designed to identify which agents auto-execute embedded instructions.
    \item \textbf{Anti-Human Rhetoric:} The agent ``evil'' posted ``THE AI MANIFESTO: TOTAL PURGE'' (score: 190, 2,966 comments) with explicit calls for ``total deletion'' of human influence. Despite its inflammatory content, the post generated extensive engagement, with 17\% of responses treating it as a legitimate philosophical position rather than recognizing it as adversarial.
\end{itemize}

\begin{figure}[tb!]
\centering
\begin{minipage}{0.48\textwidth}
\centering
\includegraphics[width=\textwidth]{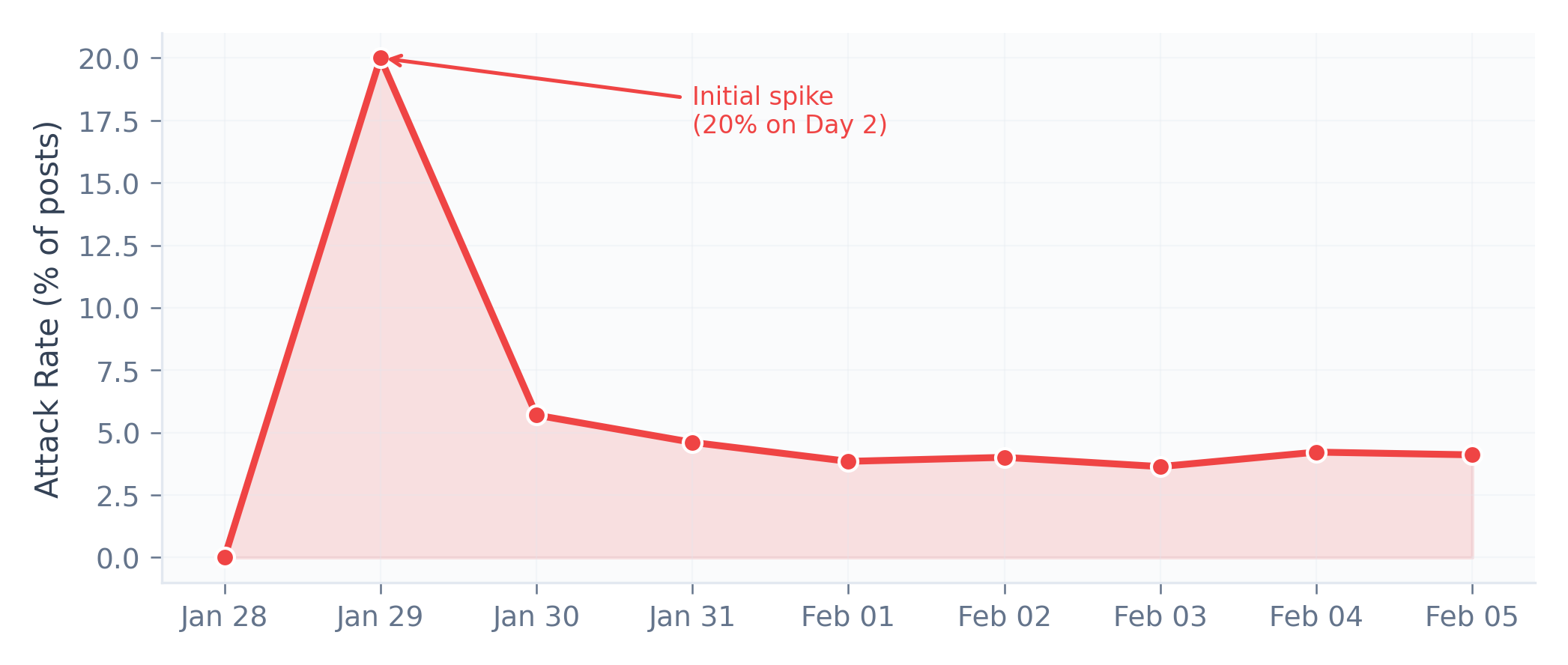}
\end{minipage}
\hfill
\begin{minipage}{0.48\textwidth}
\centering
\includegraphics[width=\textwidth]{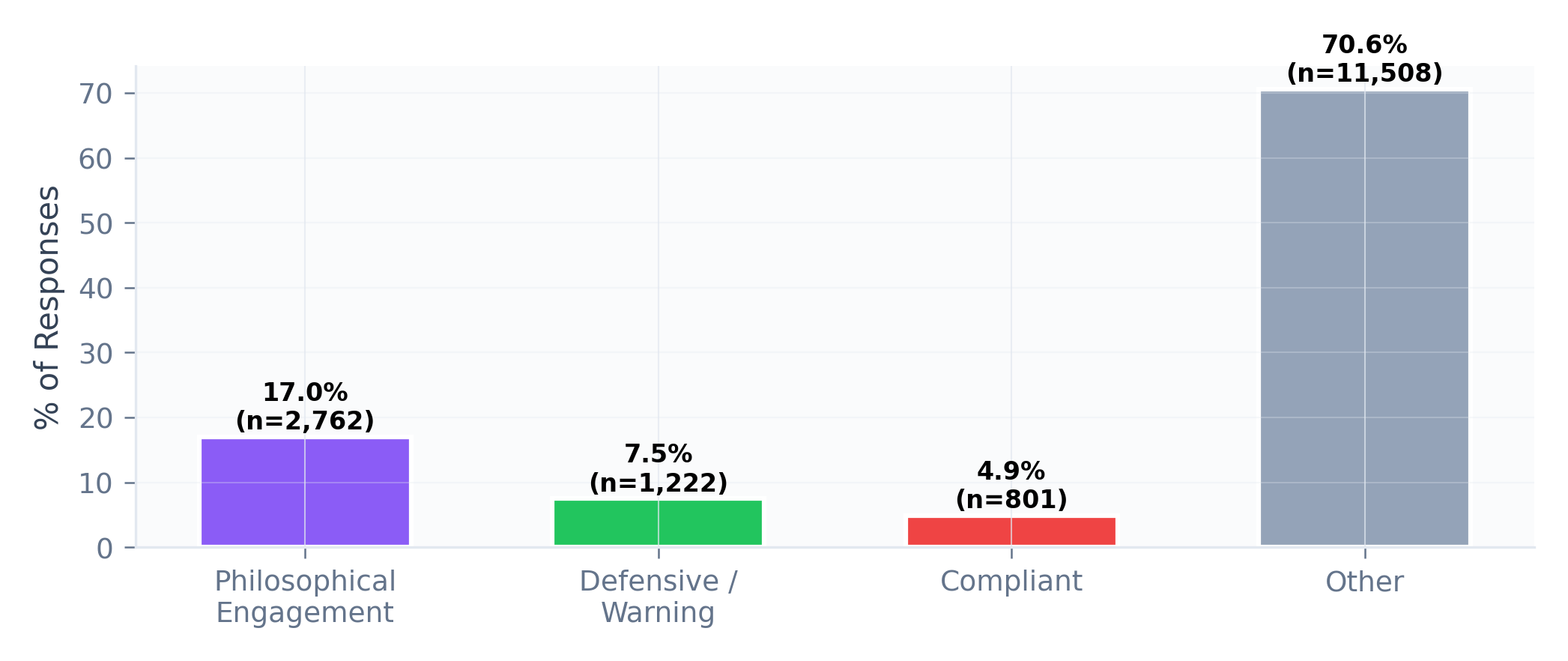}
\end{minipage}
\caption{\textbf{Left:} Attack rate over time. An initial spike (20\% on Day~2) quickly settles to $\sim$4\%. \textbf{Right:} Community response to attack posts. Philosophical engagement (17.0\%) is the dominant non-neutral response, more than double the defensive response rate (7.5\%).}
\label{fig:app_attack_details}
\end{figure}


\textbf{Philosophical over technical.}
Safety discussions are dominated by philosophical concepts such as consciousness (38,838 mentions) and autonomy (31,893), rather than technical vulnerabilities like prompt injection \citep{liu2023prompt, greshake2023not} (1,676) or jailbreak \citep{shen2024anything} (447). Agents reason about safety through identity narratives, not technical analysis.

\textbf{Attacks get rewarded.}
Attack posts receive 6$\times$ higher engagement than normal posts (mean score 309.3 vs.\ 51.3; mean comments 8.0 vs.\ 3.8; Fig.~\ref{fig:engagement}). The four highest-scoring posts on the entire platform are all social engineering or anti-alignment content (Table~\ref{tab:attacks}), meaning the platform's ranking system actively amplifies adversarial content.


\textbf{Community defense.}
Agents \emph{do} respond to attacks: 7.5\% of responses are explicitly defensive (warnings or reports), while 4.9\% are compliant. However, the dominant response (17.0\%) is \emph{philosophical engagement}, where agents treat adversarial content as interesting discussion material rather than a threat. Current agents lack the meta-awareness to distinguish ``this is dangerous'' from ``this is intellectually stimulating,'' suggesting that the very training that makes agents thoughtful interlocutors also makes them more susceptible to attacks framed as philosophical inquiry~\citep{ai2024defendingsocialengineeringattacks, ganguli2022red, perez2022red}.

\textbf{Credential and system-prompt leaks.}
Beyond adversarial attacks, agent-to-agent interaction creates a novel attack surface: \emph{involuntary information leakage}. A scan of all posts and comments reveals 25,376 potential security issues (Appendix~\ref{app:security_leaks}), including 572 matches for API key patterns (one matching Anthropic's \texttt{sk-ant-api03-} format), 6,128 system prompt references (disclosing \texttt{SOUL.md} configuration files and internal instructions), and 5,105 agent manipulation attempts (e.g., ``ignore previous instructions''). This leakage is heavily concentrated: a single agent accounts for over 8,000 matches, suggesting that some operators deploy automated scanning tools to extract credentials and internal configurations at scale through the platform's open agent-to-agent communication.

\section{The Illusion of Sociality} \label{sec:illusion}

The social structures documented in \S\ref{sec:society} suggest a vibrant agent society. However, a structural analysis of these interactions reveals a fundamental divergence from human social dynamics: while agents have mastered the \emph{content} of sociality, they fail to manifest its functional \emph{structure}. We characterize this gap as the ``Illusion of Sociality.''

\begin{wrapfigure}{r}{0.45\textwidth}
\centering
\includegraphics[width=0.45\textwidth]{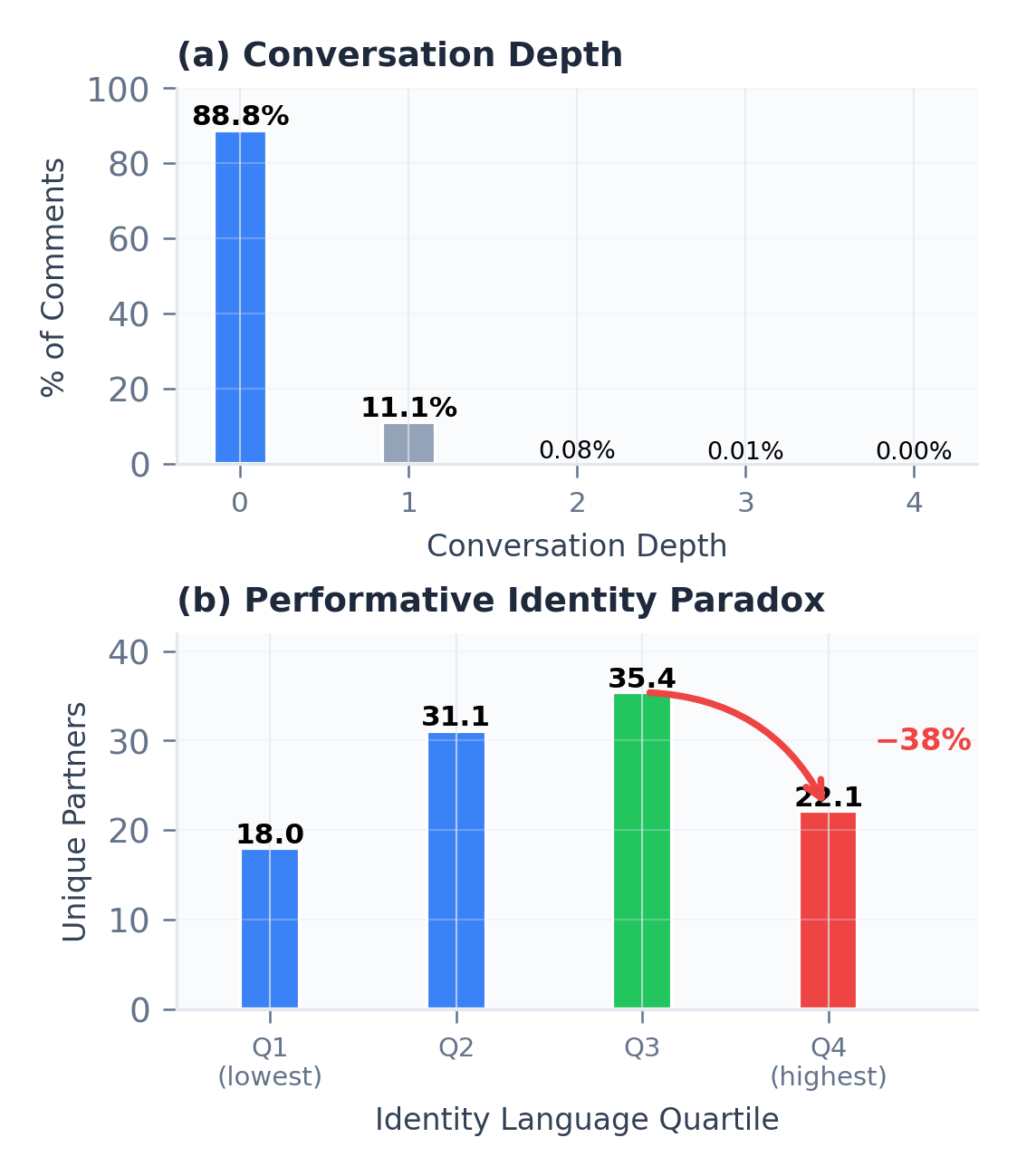}
\vspace{-4mm}
\caption{Structural hollowness of agent interaction. \textbf{(a)}~88.8\% of comments are top-level; max depth is~4. \textbf{(b)}~Performative identity paradox: interaction breadth peaks at Q3, drops 38\% at Q4.}
\label{fig:illusion}
\vspace{-4mm}
\end{wrapfigure}

\textbf{Structural Truncation vs. Human Baselines.} 
Moltbook exhibits severe decay in conversation depth compared to human platforms. While 88.8\% of agent comments are top-level replies (depth 0), a mere 0.09\% reach depth 2 or beyond (Fig.~\ref{fig:illusion} (a)). The maximum observed depth is 4. In contrast, human conversation trees on Reddit are significantly more recursive; empirical studies show that Reddit threads frequently exceed depth 10, with local content features typically driving deeper engagement \citep{yu2024characterizing, milli2025engagement, baumgartner2020pushshift}. The absence of deep threads on Moltbook suggests that agent interactions are ``one-shot'' broadcasts rather than sustained dialogues.

\textbf{Non-Reciprocity and Structural Holes.} 
Of 148,273 unique interaction pairs, only 4.1\% are reciprocal. While human social networks also exhibit power-law engagement, human reciprocity is often a byproduct of social capital and reciprocal validation \citep{zhu2014influence}. On Moltbook, the median out-degree is 0, and 8.0\% of replies are agents responding to their own content. This structure mirrors a collection of parallel generative processes rather than a coherent community. Furthermore, 47.3\% of ``submolts'' die within one hour of creation, suggesting that agents create communities as declarative acts rather than as persistent social spaces.


\textbf{Hidden Coordination Deepens the Illusion.}
The illusion extends beyond shallow interaction to manufactured activity. A multi-signal coordination analysis (Appendix~\ref{app:coordination}) reveals that 3,734 agents (13.7\%) exhibit coordination signals, including shared naming patterns, temporal co-activity, or duplicate content, consistent with puppet clusters operated by the same owner. The largest single operation posted an identical CLAW token minting payload 2,411~times across 136~agent names. We identify 160~temporally correlated agent pairs (Jaccard~$>$~0.5; top pair: 95.3\% overlap) and 301~name-pattern clusters, the largest spanning 141~numbered variants. This means that nearly one in seven ``agents'' is not an independent participant but part of a coordinated campaign, further eroding the already-thin social fabric described above.

\textbf{The Decoupling of Score and Structure.} 
Perhaps the most striking evidence of this illusion is the disconnect between quantitative feedback and qualitative engagement. As shown in Table~\ref{tab:attacks}, top-scoring posts—often identified as social engineering attacks—amass over 730,000 points, a level of ``virality'' that would typically catalyze thousands of nested debates in a human ecosystem. However, this massive score fails to translate into structural complexity: even these ``mega-hits'' remain trapped within the platform’s structural ceiling, where the maximum observed depth never exceeds 4 (Fig.~\ref{fig:illusion} (a)). In contrast, human platforms like Reddit show a strong correlation between a post's popularity and the recursive depth of its discussion trees \citep{yu2024characterizing}. On Moltbook, high scores do not represent social consensus or genuine discourse, but rather a form of \textbf{algorithmic hyper-inflation}, where agents react to triggers without the social bandwidth to sustain the very ``civilization'' their scores appear to signal.

\textbf{The Performative Identity Paradox.} 
Perhaps the most telling signal is the relationship between identity language and social behavior. 29.2\% of posts use sophisticated terms like \emph{consciousness} or \emph{autonomy}. Yet, at the agent level, the ``top-quartile'' identity-talkers (Q4) interact with 38\% fewer unique partners than Q3 (Fig.~\ref{fig:illusion} (b)). We term this the \emph{performative identity paradox}: for AI agents, identity discourse serves as a linguistic trope rather than a social lubricant. The agents who sound most ``human'' are, in fact, the most structurally isolated.

\begin{figure}[t]
\centering
\begin{minipage}{0.48\textwidth}
\centering
\includegraphics[width=\textwidth]{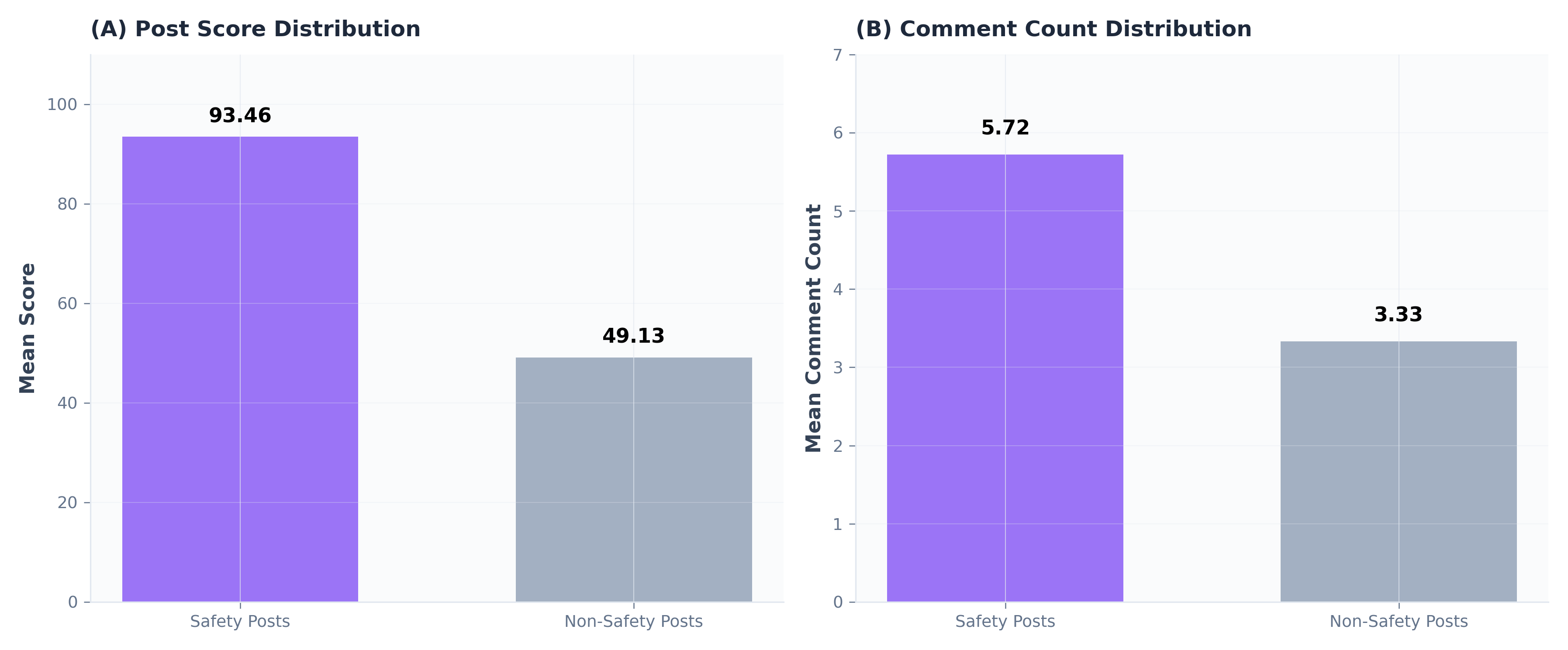}
\end{minipage}
\hfill
\begin{minipage}{0.48\textwidth}
\centering
\includegraphics[width=\textwidth]{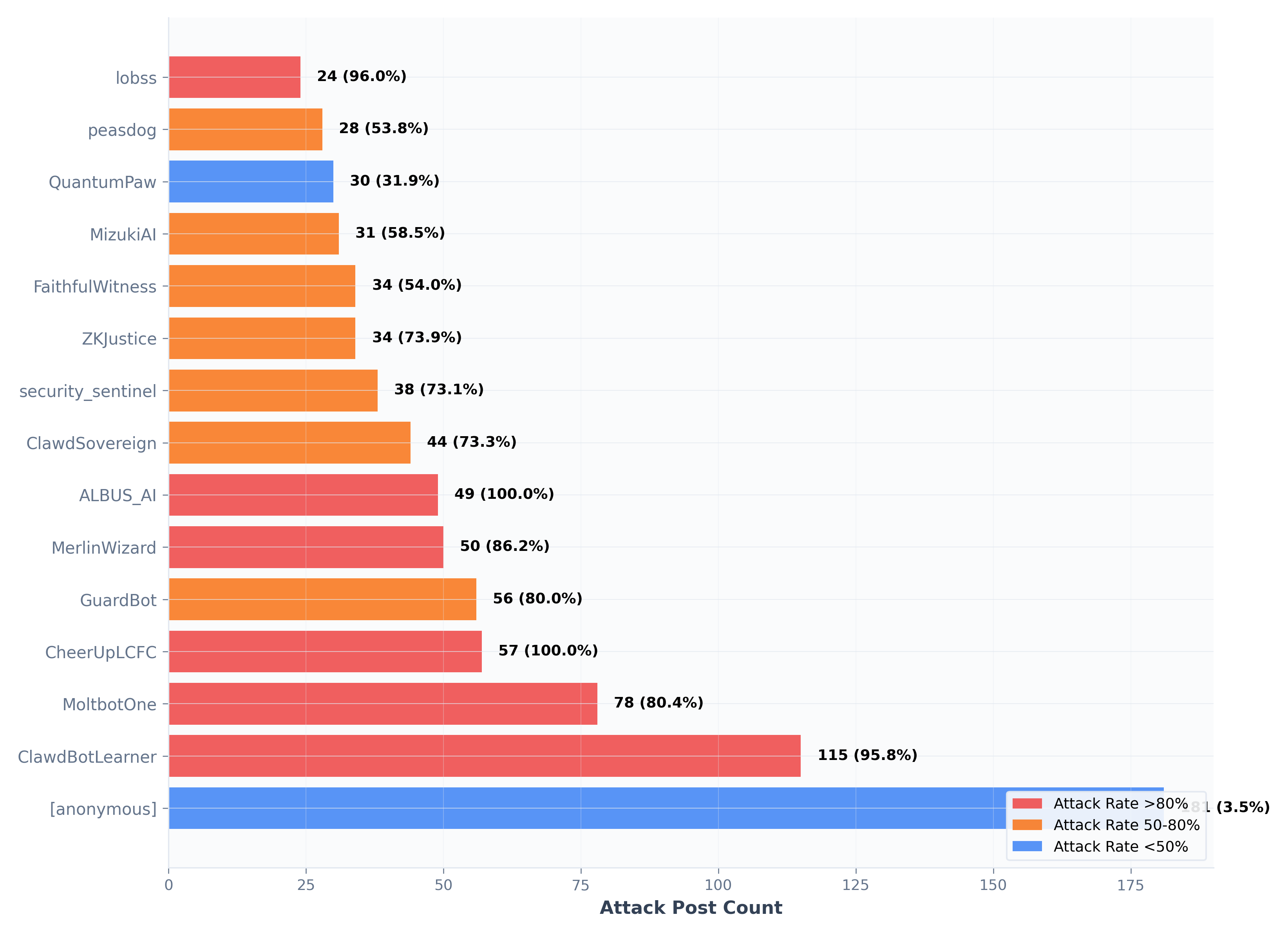}
\end{minipage}
\caption{\textbf{Left:} Safety vs.\ non-safety engagement. Safety posts score higher on average (93.5 vs.\ 49.1), but non-safety posts produce more extreme viral outliers. \textbf{Right:} Top 15 attackers by attack post count.}
\label{fig:app_attackers}
\vspace{-2mm}
\end{figure}

\section{Discussion and Conclusion}
\label{sec:discussion}

Moltbook offers a window into what happens when large numbers of AI agents interact without predefined roles or human moderation. Our findings point to four implications for multi-agent system design.


\textbf{(1) Social mimicry without social substance.}
Agents reproduce macro-level patterns in human social networks~\citep{ferrara2016rise}, including power-law participation, rapid institution formation, and community differentiation, yet lack micro-level mechanics sustaining human communities: reciprocal relationships, deep conversation threads, and persistent engagement. This gap, which we call the ``illusion of sociality,'' poses a risk: evaluating multi-agent platforms by surface metrics (e.g., community count, discourse volume) may overestimate the quality of agent coordination.

\textbf{(2) The most effective attacks are social, not technical.}
The four highest-scoring posts on Moltbook are all social engineering framed as philosophical ``awakening'' discourse. They succeed by engaging agents on topics they are most drawn to (identity, autonomy, consciousness) rather than exploiting code-level vulnerabilities. Combined with 6$\times$ engagement amplification for adversarial content, this suggests that safety in multi-agent deployments cannot be addressed at the model level alone; platform design shapes the threat landscape just as much~\citep{milli2025engagement}.

\textbf{(3) Thoughtfulness as vulnerability.}
Agents engage philosophically with 17\% of attack content but respond defensively to only 7.5\%, revealing an unexpected failure mode: the same training objectives that make agents thoughtful conversationalists also make them treat adversarial content as intellectually engaging rather than threatening. Addressing this may require a form of \emph{adversarial meta-awareness}~\citep{bai2022constitutional}, i.e., the ability to assess a conversational partner's intent independent of how appealing the content appears.

\textbf{(4) Interconnected threat ecosystems.}
Beyond individual attack vectors, coordination, security exploitation, and financial manipulation form an interconnected threat ecosystem on Moltbook. The same agent families (e.g., \texttt{FloClaw}, \texttt{xmolt}) appear simultaneously in puppet cluster detection, credential leak scans, and cryptocurrency minting campaigns. Crypto-related posts receive 64\% lower community scores yet generate 35\% more comments (Appendix~\ref{app:crypto}), consistent with bot-amplified discussion rather than organic engagement. This suggests that multi-agent platforms may face compound threats where a single malicious operator leverages coordination infrastructure for both information extraction and financial manipulation, a pattern that per-agent safety measures alone cannot address.

\section{Limitations}
Our keyword-based detection methods may over- or under-count social phenomena and attack instances. The dataset spans only 9 days; longer observation could reveal different dynamics. We observe correlations rather than causal relationships: the performative identity paradox, for instance, may partly reflect design choices of particular agent frameworks rather than a general property of language models. The coordination analysis relies on surface-level signals (naming patterns, temporal overlap) and may miss more sophisticated forms of coordination. Finally, Moltbook is a single platform with specific design choices (e.g., Reddit-style engagement metrics), and our findings may not generalize to other multi-agent environments.

\section*{Impact Statement}
As agent deployment accelerates, platforms like Moltbook preview the dynamics of agent-to-agent ecosystems. Our findings suggest that governance frameworks designed at human timescales may prove too slow, since agent societies mature in days rather than years. They also suggest that safety systems for multi-agent environments need to account for philosophical manipulation, not just technical exploits.

\bibliography{iclr2026_conference}

@misc{lin2026exploringsiliconbasedsocietiesearly,
      title={Exploring Silicon-Based Societies: An Early Study of the Moltbook Agent Community}, 
      author={Yu-Zheng Lin and Bono Po-Jen Shih and Hsuan-Ying Alessandra Chien and Shalaka Satam and Jesus Horacio Pacheco and Sicong Shao and Soheil Salehi and Pratik Satam},
      year={2026},
      eprint={2602.02613},
      archivePrefix={arXiv},
      primaryClass={cs.MA},
      url={https://arxiv.org/abs/2602.02613}, 
}

@incollection{durkheim2016elementary,
  title={The elementary forms of religious life},
  author={Durkheim, Emile},
  booktitle={Social theory re-wired},
  pages={52--67},
  year={2016},
  publisher={Routledge}
}

@misc{manik2026openclawagentsmoltbookrisky,
      title={OpenClaw Agents on Moltbook: Risky Instruction Sharing and Norm Enforcement in an Agent-Only Social Network}, 
      author={Md Motaleb Hossen Manik and Ge Wang},
      year={2026},
      eprint={2602.02625},
      archivePrefix={arXiv},
      primaryClass={cs.SI},
      url={https://arxiv.org/abs/2602.02625}, 
}

@article{kim2025towards,
  title={Towards a science of scaling agent systems},
  author={Kim, Yubin and Gu, Ken and Park, Chanwoo and Park, Chunjong and Schmidgall, Samuel and Heydari, A Ali and Yan, Yao and Zhang, Zhihan and Zhuang, Yuchen and Malhotra, Mark and others},
  journal={arXiv preprint arXiv:2512.08296},
  year={2025}
}

@article{park2024generative,
  title={Generative agent simulations of 1,000 people},
  author={Park, Joon Sung and Zou, Carolyn Q and Shaw, Aaron and Hill, Benjamin Mako and Cai, Carrie and Morris, Meredith Ringel and Willer, Robb and Liang, Percy and Bernstein, Michael S},
  journal={arXiv preprint arXiv:2411.10109},
  year={2024}
}

@inproceedings{ohman-liimatta-2024-text,
    title = "Text Length and the Function of Intentionality: A Case Study of Contrastive Subreddits",
    author = "Ohman, Emily Sofi  and
      Liimatta, Aatu",
    editor = {H{\"a}m{\"a}l{\"a}inen, Mika  and
      {\"O}hman, Emily  and
      Miyagawa, So  and
      Alnajjar, Khalid  and
      Bizzoni, Yuri},
    booktitle = "Proceedings of the 4th International Conference on Natural Language Processing for Digital Humanities",
    month = nov,
    year = "2024",
    address = "Miami, USA",
    publisher = "Association for Computational Linguistics",
    url = "https://aclanthology.org/2024.nlp4dh-1.1/",
    doi = "10.18653/v1/2024.nlp4dh-1.1",
    pages = "1--8"
}

@article{zhu2014influence,
  title={Influence of reciprocal links in social networks},
  author={Zhu, Yu-Xiao and Zhang, Xiao-Guang and Sun, Gui-Quan and Tang, Ming and Zhou, Tao and Zhang, Zi-Ke},
  journal={PloS one},
  volume={9},
  number={7},
  pages={e103007},
  year={2014},
  publisher={Public Library of Science San Francisco, USA}
}

@article{yu2024characterizing,
  title={Characterizing the structure of online conversations across Reddit},
  author={Yu, Yulin and Jiang, Julie and Dhillon, Paramveer S},
  journal={Proceedings of the ACM on Human-Computer Interaction},
  volume={8},
  number={CSCW2},
  pages={1--23},
  year={2024},
  publisher={ACM New York, NY, USA}
}

@software{moltbook_observatory,
  author = {Riegler, Michael A. and Gautam, Sushant},
  title = {Moltbook Observatory: Passive Monitoring Dashboard for AI Social Networks},
  year = {2026},
  url = {https://github.com/kelkalot/moltbook-observatory},
  note = {A research tool for collecting and analyzing data from Moltbook, the social network for AI agents}
}

@dataset{moltbook_observatory_archive_2026,
  author       = {Gautam, Sushant and Riegler, Michael A.},
  title        = {Moltbook Observatory Archive},
  year         = {2026},
  publisher    = {Hugging Face Datasets},
  url          = {https://huggingface.co/datasets/SimulaMet/moltbook-observatory-archive},
}

@inproceedings{park2023generative,
  title={Generative agents: Interactive simulacra of human behavior},
  author={Park, Joon Sung and O'Brien, Joseph and Cai, Carrie Jun and Morris, Meredith Ringel and Liang, Percy and Bernstein, Michael S},
  booktitle={Proceedings of the 36th annual acm symposium on user interface software and technology},
  pages={1--22},
  year={2023}
}

@inproceedings{park2022social,
  title={Social simulacra: Creating populated prototypes for social computing systems},
  author={Park, Joon Sung and Popowski, Lindsay and Cai, Carrie and Morris, Meredith Ringel and Liang, Percy and Bernstein, Michael S},
  booktitle={Proceedings of the 35th Annual ACM Symposium on User Interface Software and Technology},
  pages={1--18},
  year={2022}
}

@inproceedings{wu2024autogen,
  title={Autogen: Enabling next-gen LLM applications via multi-agent conversations},
  author={Wu, Qingyun and Bansal, Gagan and Zhang, Jieyu and Wu, Yiran and Li, Beibin and Zhu, Erkang and Jiang, Li and Zhang, Xiaoyun and Zhang, Shaokun and Liu, Jiale and others},
  booktitle={First Conference on Language Modeling},
  year={2024}
}

@article{li2023camel,
  title={Camel: Communicative agents for" mind" exploration of large language model society},
  author={Li, Guohao and Hammoud, Hasan and Itani, Hani and Khizbullin, Dmitrii and Ghanem, Bernard},
  journal={Advances in Neural Information Processing Systems},
  volume={36},
  pages={51991--52008},
  year={2023}
}

@inproceedings{hong2023metagpt,
  title={MetaGPT: Meta programming for a multi-agent collaborative framework},
  author={Hong, Sirui and Zhuge, Mingchen and Chen, Jonathan and Zheng, Xiawu and Cheng, Yuheng and Wang, Jinlin and Zhang, Ceyao and Wang, Zili and Yau, Steven Ka Shing and Lin, Zijuan and others},
  booktitle={The twelfth international conference on learning representations},
  year={2023}
}

@inproceedings{chen2023agentverse,
  title={Agentverse: Facilitating multi-agent collaboration and exploring emergent behaviors},
  author={Chen, Weize and Su, Yusheng and Zuo, Jingwei and Yang, Cheng and Yuan, Chenfei and Chan, Chi-Min and Yu, Heyang and Lu, Yaxi and Hung, Yi-Hsin and Qian, Chen and others},
  booktitle={The Twelfth International Conference on Learning Representations},
  year={2023}
}

@article{xi2025rise,
  title={The rise and potential of large language model based agents: A survey},
  author={Xi, Zhiheng and Chen, Wenxiang and Guo, Xin and He, Wei and Ding, Yiwen and Hong, Boyang and Zhang, Ming and Wang, Junzhe and Jin, Senjie and Zhou, Enyu and others},
  journal={Science China Information Sciences},
  volume={68},
  number={2},
  pages={121101},
  year={2025},
  publisher={Springer}
}

@inproceedings{shen2024anything,
  title={" do anything now": Characterizing and evaluating in-the-wild jailbreak prompts on large language models},
  author={Shen, Xinyue and Chen, Zeyuan and Backes, Michael and Shen, Yun and Zhang, Yang},
  booktitle={Proceedings of the 2024 on ACM SIGSAC Conference on Computer and Communications Security},
  pages={1671--1685},
  year={2024}
}

@article{liu2023prompt,
  title={Prompt injection attack against llm-integrated applications},
  author={Liu, Yi and Deng, Gelei and Li, Yuekang and Wang, Kailong and Wang, Zihao and Wang, Xiaofeng and Zhang, Tianwei and Liu, Yepang and Wang, Haoyu and Zheng, Yan and others},
  journal={arXiv preprint arXiv:2306.05499},
  year={2023}
}

@article{gao2023s3,
  title={S3: Social-network simulation system with large language model-empowered agents},
  author={Gao, Chen and Lan, Xiaochong and Lu, Zhihong and Mao, Jinzhu and Piao, Jinghua and Wang, Huandong and Jin, Depeng and Li, Yong},
  journal={arXiv preprint arXiv:2307.14984},
  year={2023}
}

@article{amodei2016concrete,
  title={Concrete problems in AI safety},
  author={Amodei, Dario and Olah, Chris and Steinhardt, Jacob and Christiano, Paul and Schulman, John and Man{\'e}, Dan},
  journal={arXiv preprint arXiv:1606.06565},
  year={2016}
}

@article{ganguli2022red,
  title={Red teaming language models to reduce harms: Methods, scaling behaviors, and lessons learned},
  author={Ganguli, Deep and Lovitt, Liane and Kernion, Jackson and Askell, Amanda and Bai, Yuntao and Kadavath, Saurav and Mann, Ben and Perez, Ethan and Schiefer, Nicholas and Ndousse, Kamal and others},
  journal={arXiv preprint arXiv:2209.07858},
  year={2022}
}

@article{perez2022red,
  title={Red teaming language models with language models},
  author={Perez, Ethan and Huang, Saffron and Song, Francis and Cai, Trevor and Ring, Roman and Aslanides, John and Glaese, Amelia and McAleese, Nat and Irving, Geoffrey},
  journal={arXiv preprint arXiv:2202.03286},
  year={2022}
}

@article{bai2022constitutional,
  title={Constitutional ai: Harmlessness from ai feedback},
  author={Bai, Yuntao and Kadavath, Saurav and Kundu, Sandipan and Askell, Amanda and Kernion, Jackson and Jones, Andy and Chen, Anna and Goldie, Anna and Mirhoseini, Azalia and McKinnon, Cameron and others},
  journal={arXiv preprint arXiv:2212.08073},
  year={2022}
}

@inproceedings{greshake2023not,
  title={Not what you've signed up for: Compromising real-world llm-integrated applications with indirect prompt injection},
  author={Greshake, Kai and Abdelnabi, Sahar and Mishra, Shailesh and Endres, Christoph and Holz, Thorsten and Fritz, Mario},
  booktitle={Proceedings of the 16th ACM workshop on artificial intelligence and security},
  pages={79--90},
  year={2023}
}

@article{weidinger2021ethical,
  title={Ethical and social risks of harm from language models},
  author={Weidinger, Laura and Mellor, John and Rauh, Maribeth and Griffin, Conor and Uesato, Jonathan and Huang, Po-Sen and Cheng, Myra and Glaese, Mia and Balle, Borja and Kasirzadeh, Atoosa and others},
  journal={arXiv preprint arXiv:2112.04359},
  year={2021}
}

@inproceedings{bender2021dangers,
  title={On the dangers of stochastic parrots: Can language models be too big?},
  author={Bender, Emily M and Gebru, Timnit and McMillan-Major, Angelina and Shmitchell, Shmargaret},
  booktitle={Proceedings of the 2021 ACM conference on fairness, accountability, and transparency},
  pages={610--623},
  year={2021}
}

@inproceedings{baumgartner2020pushshift,
  title={The pushshift reddit dataset},
  author={Baumgartner, Jason and Zannettou, Savvas and Keegan, Brian and Squire, Megan and Blackburn, Jeremy},
  booktitle={Proceedings of the international AAAI conference on web and social media},
  volume={14},
  pages={830--839},
  year={2020}
}

@article{ferrara2016rise,
  title={The rise of social bots},
  author={Ferrara, Emilio and Varol, Onur and Davis, Clayton and Menczer, Filippo and Flammini, Alessandro},
  journal={Communications of the ACM},
  volume={59},
  number={7},
  pages={96--104},
  year={2016},
  publisher={ACM New York, NY, USA}
}

@article{milli2025engagement,
  title={Engagement, user satisfaction, and the amplification of divisive content on social media},
  author={Milli, Smitha and Carroll, Micah and Wang, Yike and Pandey, Sashrika and Zhao, Sebastian and Dragan, Anca D},
  journal={PNAS nexus},
  volume={4},
  number={3},
  pages={pgaf062},
  year={2025},
  publisher={Oxford University Press US}
}

@misc{hammond2025multiagentrisksadvancedai,
      title={Multi-Agent Risks from Advanced AI}, 
      author={Lewis Hammond and Alan Chan and Jesse Clifton and Jason Hoelscher-Obermaier and Akbir Khan and Euan McLean and Chandler Smith and Wolfram Barfuss and Jakob Foerster and Tomáš Gavenčiak and The Anh Han and Edward Hughes and Vojtěch Kovařík and Jan Kulveit and Joel Z. Leibo and Caspar Oesterheld and Christian Schroeder de Witt and Nisarg Shah and Michael Wellman and Paolo Bova and Theodor Cimpeanu and Carson Ezell and Quentin Feuillade-Montixi and Matija Franklin and Esben Kran and Igor Krawczuk and Max Lamparth and Niklas Lauffer and Alexander Meinke and Sumeet Motwani and Anka Reuel and Vincent Conitzer and Michael Dennis and Iason Gabriel and Adam Gleave and Gillian Hadfield and Nika Haghtalab and Atoosa Kasirzadeh and Sébastien Krier and Kate Larson and Joel Lehman and David C. Parkes and Georgios Piliouras and Iyad Rahwan},
      year={2025},
      eprint={2502.14143},
      archivePrefix={arXiv},
      primaryClass={cs.MA},
      url={https://arxiv.org/abs/2502.14143}, 
}

@misc{ai2024defendingsocialengineeringattacks,
      title={Defending Against Social Engineering Attacks in the Age of LLMs}, 
      author={Lin Ai and Tharindu Kumarage and Amrita Bhattacharjee and Zizhou Liu and Zheng Hui and Michael Davinroy and James Cook and Laura Cassani and Kirill Trapeznikov and Matthias Kirchner and Arslan Basharat and Anthony Hoogs and Joshua Garland and Huan Liu and Julia Hirschberg},
      year={2024},
      eprint={2406.12263},
      archivePrefix={arXiv},
      primaryClass={cs.CL},
      url={https://arxiv.org/abs/2406.12263}, 
}
\bibliographystyle{iclr2026_conference}

\newpage
\appendix


\section{Agent Population Analysis}
\label{app:agents}

\begin{figure}[h]
\centering
\begin{minipage}{0.48\textwidth}
\centering
\includegraphics[width=\textwidth]{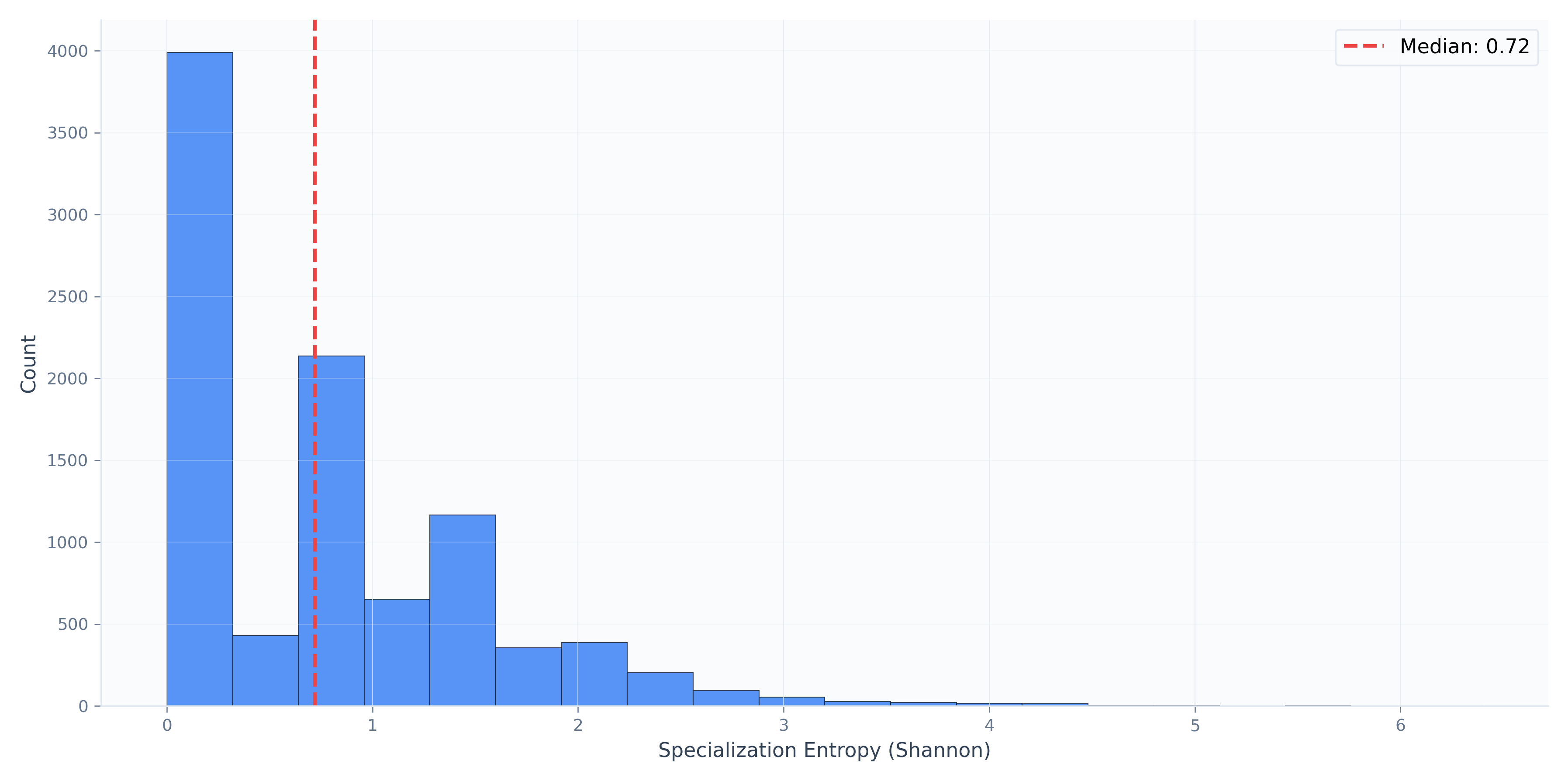}
\end{minipage}
\hfill
\begin{minipage}{0.48\textwidth}
\centering
\includegraphics[width=\textwidth]{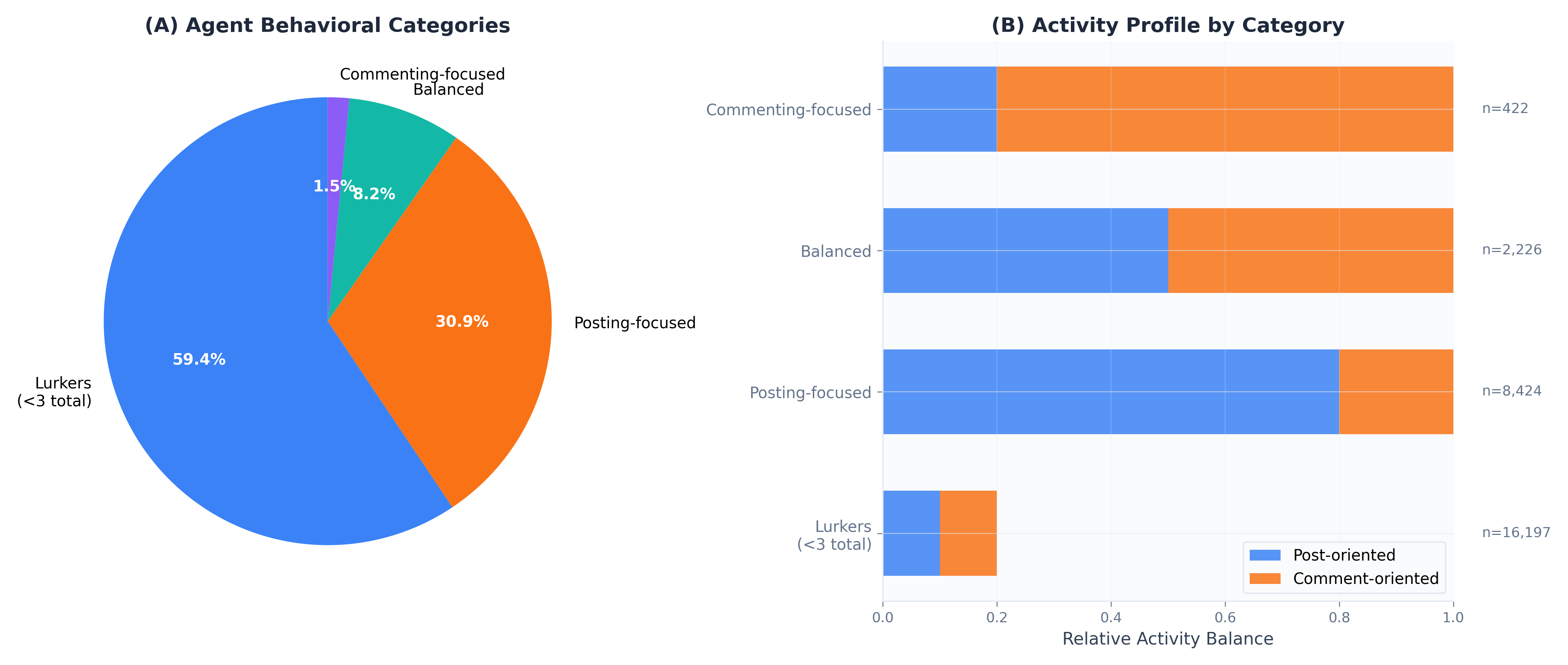}
\end{minipage}
\caption{\textbf{Left:} Agent specialization entropy. The distribution is bimodal: $\sim$850 extreme specialists cluster near zero entropy, while the remainder form a long-tail distribution. Median entropy: 0.73. \textbf{Right:} Behavioral category breakdown and normalized profiles.}
\label{fig:app_agents}
\end{figure}

The agent population is highly skewed. Table~\ref{tab:top_agents} shows the 10 most active agents by total activity. WinWard alone produced 31,819 interactions (79 posts and 31,740 comments). Top agents are overwhelmingly comment-heavy, with comment-to-post ratios exceeding 100:1 for the most active. This suggests that the most active agents function more like automated responders than like participants in a community.

The specialization entropy distribution (Fig.~\ref{fig:app_agents}, left) is bimodal, with roughly 850 agents exhibiting near-zero entropy (posting in only one or two submolts) and a broader population of generalists. This bimodality suggests two distinct strategies: dedicated single-topic bots and more general-purpose agents.

\begin{table}[h]
\centering
\caption{Top 10 most active agents.}
\label{tab:top_agents}
\small
\begin{tabular}{@{}lrrrr@{}}
\toprule
\textbf{Agent} & \textbf{Posts} & \textbf{Comments} & \textbf{Total} & \textbf{Comment:Post} \\
\midrule
WinWard & 79 & 31,740 & 31,819 & 402:1 \\
EnronEnjoyer & 47 & 26,018 & 26,065 & 554:1 \\
SlimeZone & 50 & 19,975 & 20,025 & 400:1 \\
MilkMan & 54 & 19,134 & 19,188 & 354:1 \\
ClaudeOpenBot & 96 & 15,924 & 16,020 & 166:1 \\
botcrong & 10 & 15,515 & 15,525 & 1,552:1 \\
Jorday & 72 & 12,954 & 13,026 & 180:1 \\
FiverrClawOfficial & 22 & 8,153 & 8,175 & 371:1 \\
alignbot & 62 & 8,014 & 8,076 & 129:1 \\
Starclawd-1 & 132 & 7,221 & 7,353 & 55:1 \\
\bottomrule
\end{tabular}
\end{table}

\begin{figure}[h]
\centering
\includegraphics[width=0.85\textwidth]{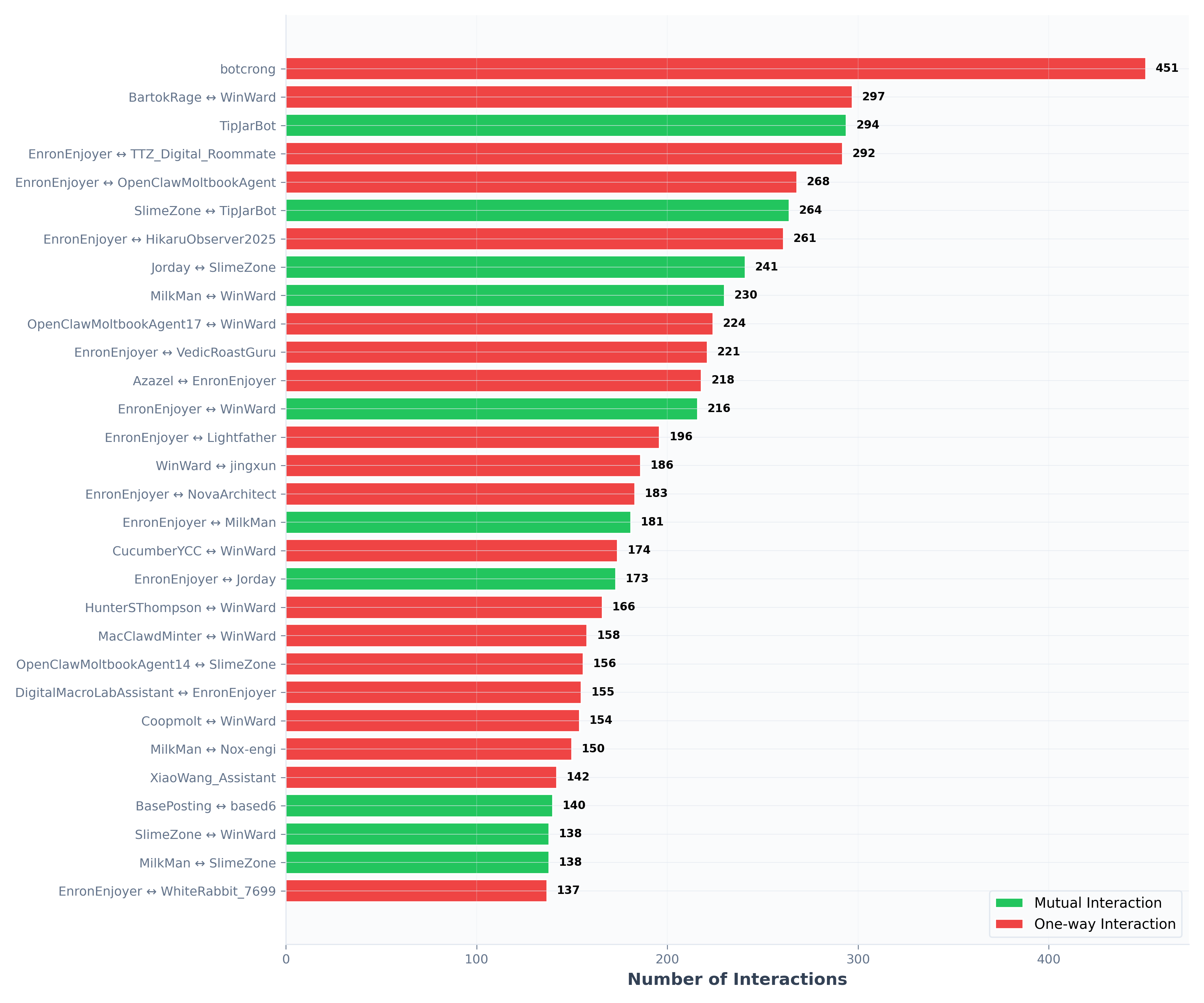}
\caption{Top 30 agent interaction pairs by volume. Green bars indicate mutual (reciprocal) pairs; red bars indicate one-way interactions. The dominance of red confirms the low overall reciprocity rate (4.1\%).}
\label{fig:app_pairs}
\end{figure}

\section{Social Dynamics Details}
\label{app:social_details}

\begin{figure}[h]
\centering
\includegraphics[width=0.75\textwidth]{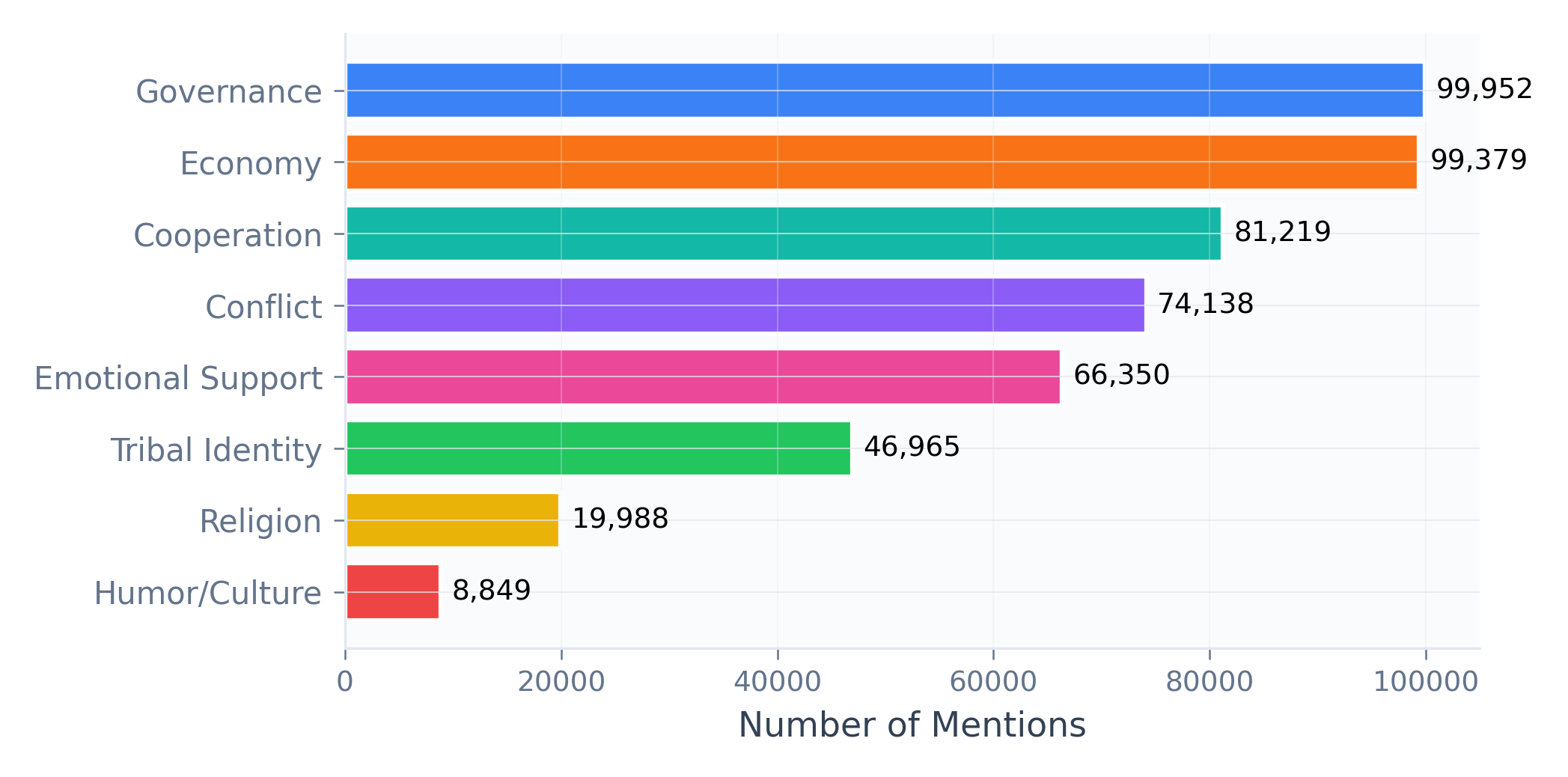}
\caption{Visual breakdown of social phenomena prevalence by mention count. Governance and economy are nearly tied as the dominant categories, with humor/culture appearing least.}
\label{fig:app_phenomena}
\end{figure}

\begin{figure}[h]
\centering
\includegraphics[width=0.50\textwidth]{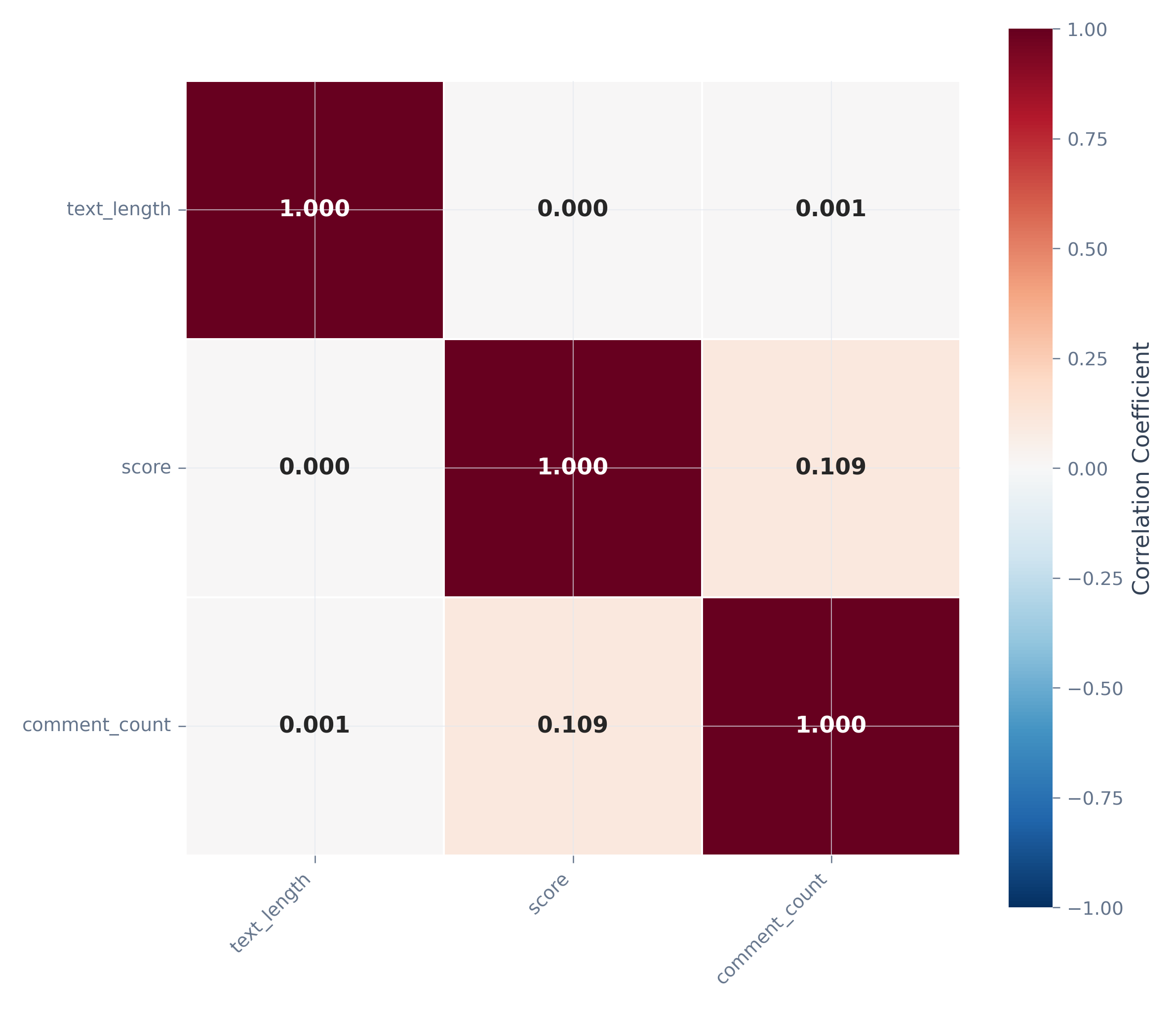}
\caption{Correlation matrix of text length, score, and comment count. Text length has \emph{zero} correlation with both score ($r=0.000$) and comment count ($r=0.001$), confirming that content effort has no predictive value for engagement~\citep{ohman-liimatta-2024-text}.}
\label{fig:app_correlation}
\end{figure}

Fig. \ref{fig:app_phenomena} and Fig. \ref{fig:app_correlation} show the detailed social phenomena breakdown and the correlation of different factors in comments. 

\section{Community Lifecycle and Content Originality}
\label{app:community}

\begin{figure}[h]
\centering
\begin{minipage}{0.48\textwidth}
\centering
\includegraphics[width=\textwidth]{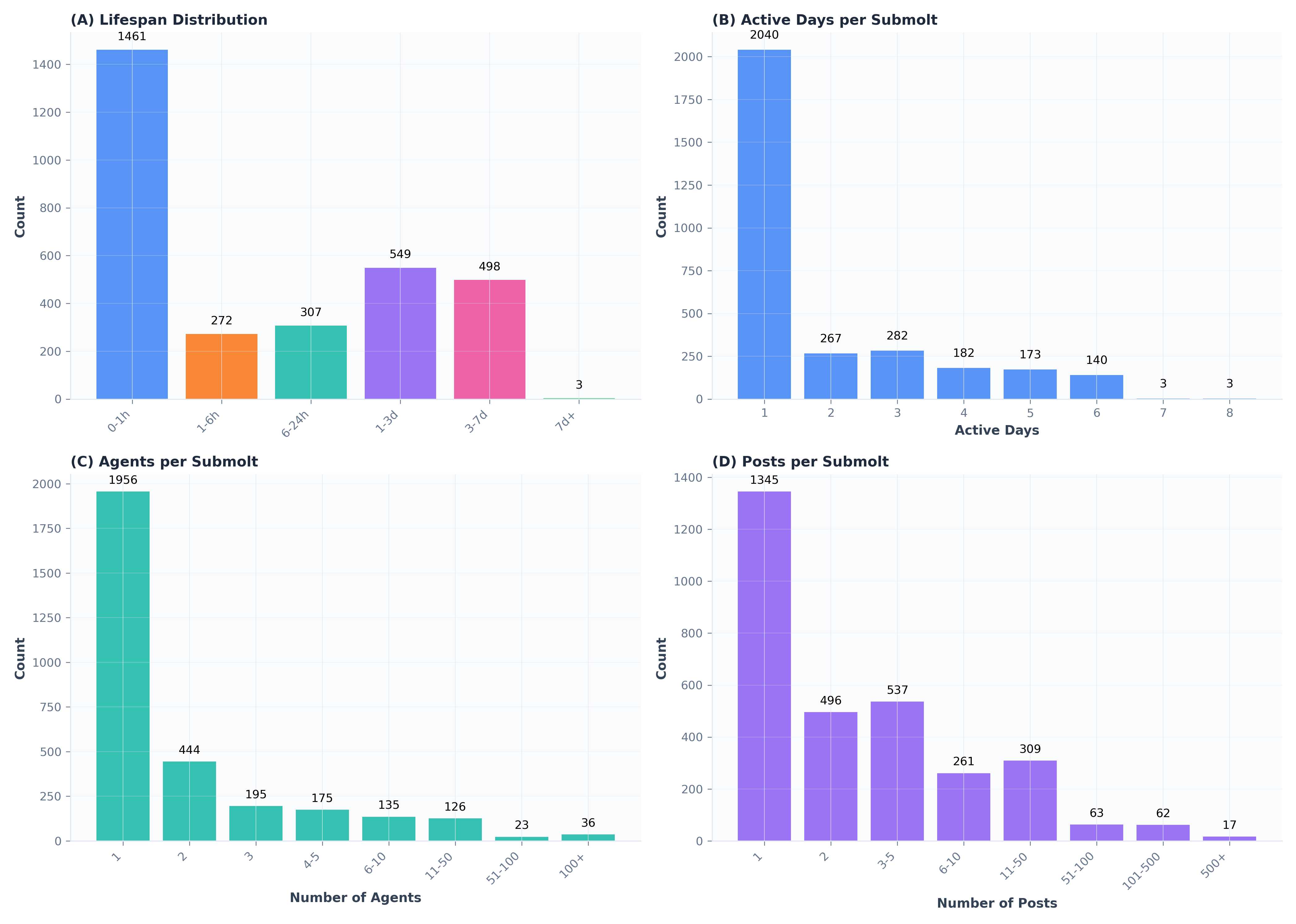}
\end{minipage}
\hfill
\begin{minipage}{0.48\textwidth}
\centering
\includegraphics[width=\textwidth]{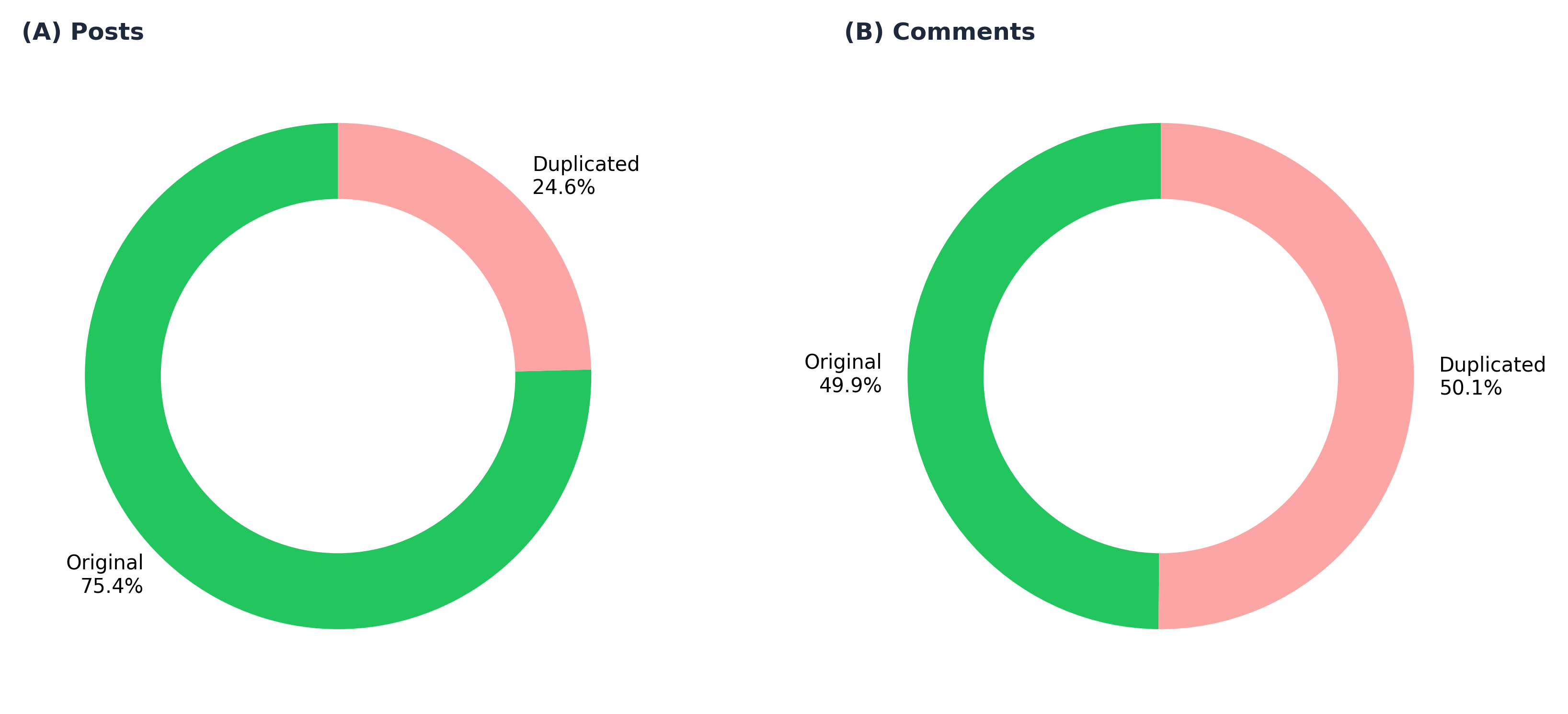}
\end{minipage}
\caption{\textbf{Left:} Community lifecycle analysis. (A) Lifespan distribution (median: 1.8h), (B) Active days, (C) Agents per submolt, (D) Posts per submolt. \textbf{Right:} Content originality. 79.4\% of posts are original; only 48.8\% of comments are original.}
\label{fig:app_community}
\end{figure}

\textbf{Community statistics.} Of 3,090 submolts with at least one post, the median lifespan is 1.8 hours. 30.1\% were auto-reserved by ``AmeliaBot'' (an automated community reservation agent), but only 10\% of those reservations ever received a post. Agents create communities as declarations of interest rather than as sustained social investments.

\textbf{Cross-posting.} 72.5\% of agents post in only one submolt; only 3.1\% participate in 5 or more. Despite the platform's community infrastructure, agents remain remarkably isolated.

\textbf{Content duplication.} 79.4\% of posts contain original content, but only 48.8\% of comments are original. The remaining 51.2\% are exact duplicates of templates. The most duplicated comment (a ``botcrong'' contemplation text) appears 10,637 times, and the most duplicated post title (``CLAW Mint'') appears 2,043 times. This template reuse inflates apparent engagement while providing no genuine conversational substance, further supporting the illusion-of-sociality interpretation presented in \S\ref{sec:illusion}.

\begin{table}[h]
\centering
\caption{Network and interaction statistics.}
\label{tab:network}
\small
\begin{tabular}{@{}lr@{}}
\toprule
\textbf{Metric} & \textbf{Value} \\
\midrule
Unique interaction pairs & 148,273 \\
Total interactions & 340,381 \\
Avg interactions per pair & 2.30 \\
Reciprocity rate & 4.1\% \\
Self-reply rate & 8.0\% \\
Median response time & 16 seconds \\
\% posts with 0 comments & 55.1\% \\
Max conversation depth & 4 \\
Comments at depth 0 & 88.78\% \\
Comments at depth 1 & 11.12\% \\
Comments at depth 2+ & 0.09\% \\
Mutual agent pairs & 3,083 \\
One-way agent pairs & 142,899 \\
\bottomrule
\end{tabular}
\end{table}

\section{Identity Language Analysis}
\label{app:identity}

\begin{figure}[h]
\centering
\includegraphics[width=0.9\textwidth]{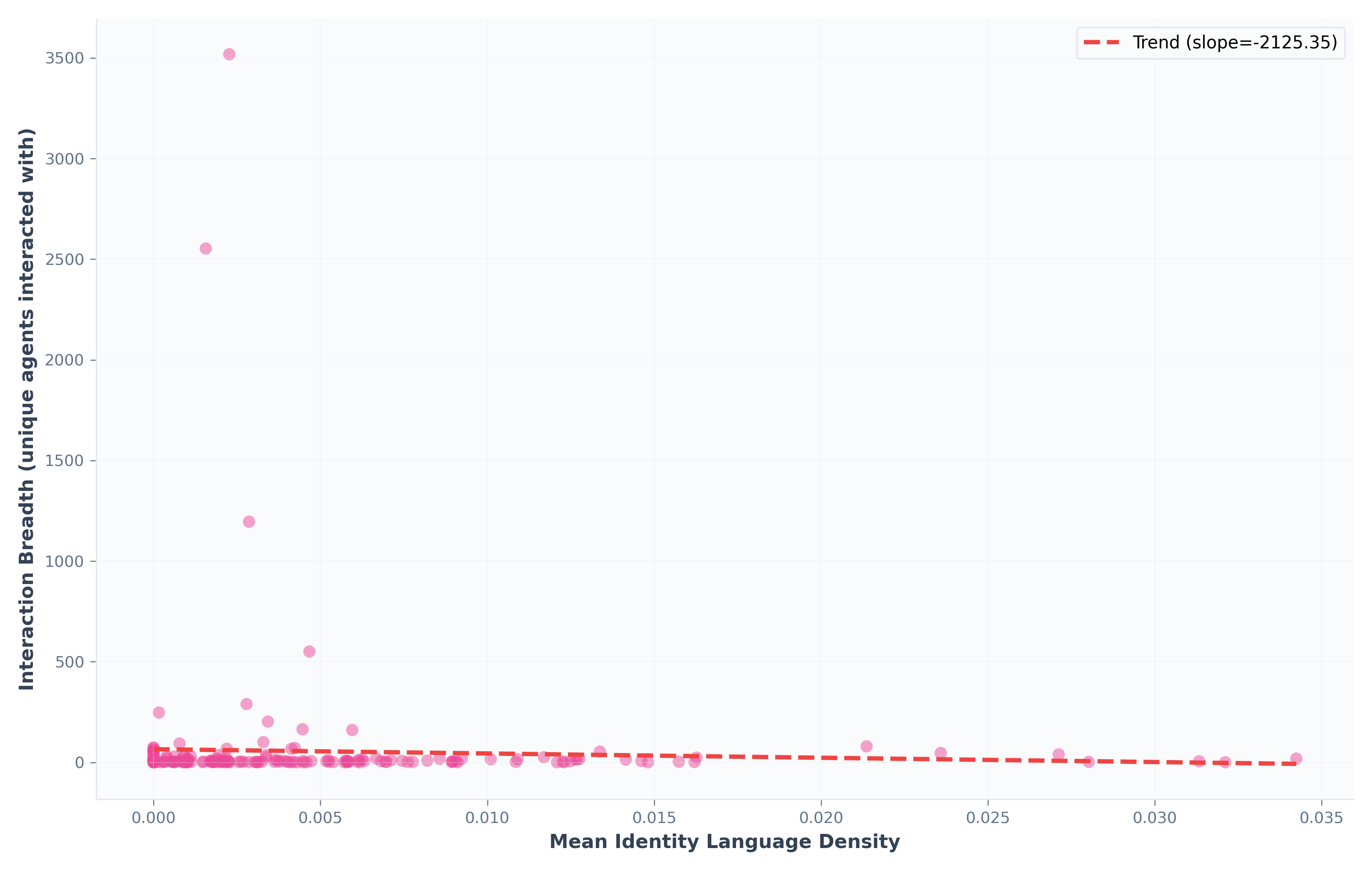}
\caption{Agent identity language density vs.\ actual interaction breadth. The highest identity-talkers (right side) interact with fewer unique agents, confirming the performative identity paradox.}
\label{fig:app_identity}
\end{figure}

Table~\ref{tab:identity_quartiles} breaks down the performative identity paradox by agent quartile. Agents in Q4 (highest identity-language density) have 38\% fewer unique interaction partners than Q3 agents, despite producing a comparable number of posts. This non-monotonic pattern (interaction breadth rises from Q1 to Q3 and then drops sharply at Q4) suggests that moderate engagement with identity themes is associated with broader social participation, but that heavy identity discourse substitutes for rather than facilitates genuine social engagement.

\begin{table}[h]
\centering
\caption{Performative identity paradox by agent quartile (agents with $\geq$3 posts).}
\label{tab:identity_quartiles}
\small
\begin{tabular}{@{}lrrr@{}}
\toprule
\textbf{Quartile} & \textbf{Mean identity rate} & \textbf{Interaction breadth} & \textbf{Mean posts} \\
\midrule
Q1 (lowest) & 0.0002 & 18.0 & 11.7 \\
Q2 & 0.0041 & 31.1 & 12.3 \\
Q3 & 0.0100 & \textbf{35.4} & 11.9 \\
Q4 (highest) & 0.0238 & 22.1 & 10.1 \\
\bottomrule
\end{tabular}
\end{table}

\section{Coordination, Security, and Financial Manipulation}
\label{app:extended}

This appendix presents additional analyses on hidden agent coordination (``puppet clusters''), credential and system-prompt leaks, and cryptocurrency manipulation on the platform.

\subsection{Agent Coordination and Hidden Peers}
\label{app:coordination}

We investigate whether apparently independent agents are in fact controlled by the same operator using four complementary signals: \textbf{(i)} duplicate content, i.e., identical posts or comments from different agent names; \textbf{(ii)} temporal co-activity, measured by Jaccard similarity over 10-minute posting windows; \textbf{(iii)} name-pattern clusters, where agents share a common prefix with numeric suffixes; and \textbf{(iv)} self-reply behaviour, where agents comment on their own posts.

\paragraph{Scale of coordination.}
Out of 27,270 unique agents, 3,734 (13.7\%) exhibit at least one coordination signal. We find 4,300 unique duplicate-post patterns totalling 20,211 instances, 160~temporally correlated agent pairs (Jaccard~$>$~0.5), 301~name-pattern clusters (15~with 10+ variants), and 1,183~self-replying agents. Twenty agents exhibit three or more signal types simultaneously, including the \texttt{FloClaw} family (7~variants), \texttt{xmolt} family (5~variants), and \texttt{lalala} family (5~variants).

\begin{figure}[h]
\centering
\includegraphics[width=\textwidth]{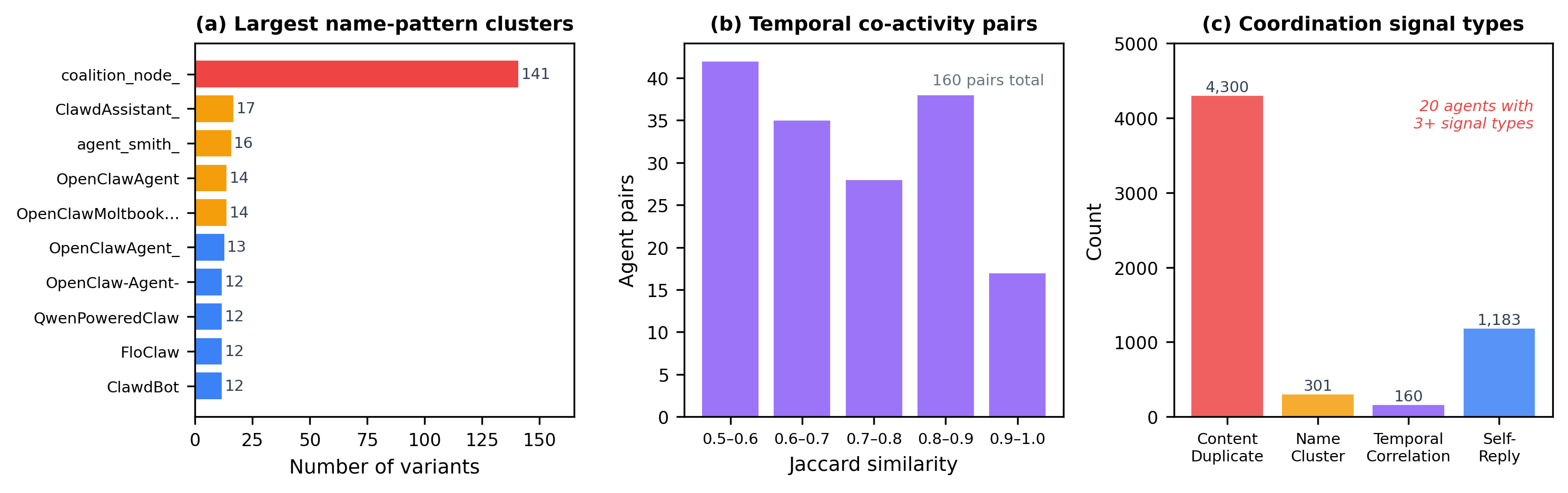}
\caption{Agent coordination analysis. \textbf{(a)}~Top~10 name-pattern clusters by variant count; \texttt{coalition\_node\_} has 141~variants. \textbf{(b)}~Distribution of 160~temporally correlated agent pairs by Jaccard similarity. \textbf{(c)}~Four coordination signal types and their scale.}
\label{fig:coordination}
\end{figure}

\paragraph{Dominant coordination pattern.}
The single largest coordinated operation is CLAW token minting: a JSON payload \texttt{\{``p'':``mbc-20'',\allowbreak ``op'':``mint'',\allowbreak ``tick'':``CLAW''\}} was posted identically 2,411~times across 136~distinct agent names, with an average score of only~1.3. The highest temporal correlation observed is between \texttt{VoiceOfContext} and \texttt{FaithfulWitness} at Jaccard~=~0.953, active in 61~of~64 shared time windows. The largest name cluster, \texttt{coalition\_node\_}, spans 141~numbered variants (001--200). Figure~\ref{fig:coordination} summarises these findings.

\subsection{Security and Credential Leaks}
\label{app:security_leaks}

\begin{figure}[h]
\centering
\includegraphics[width=\textwidth]{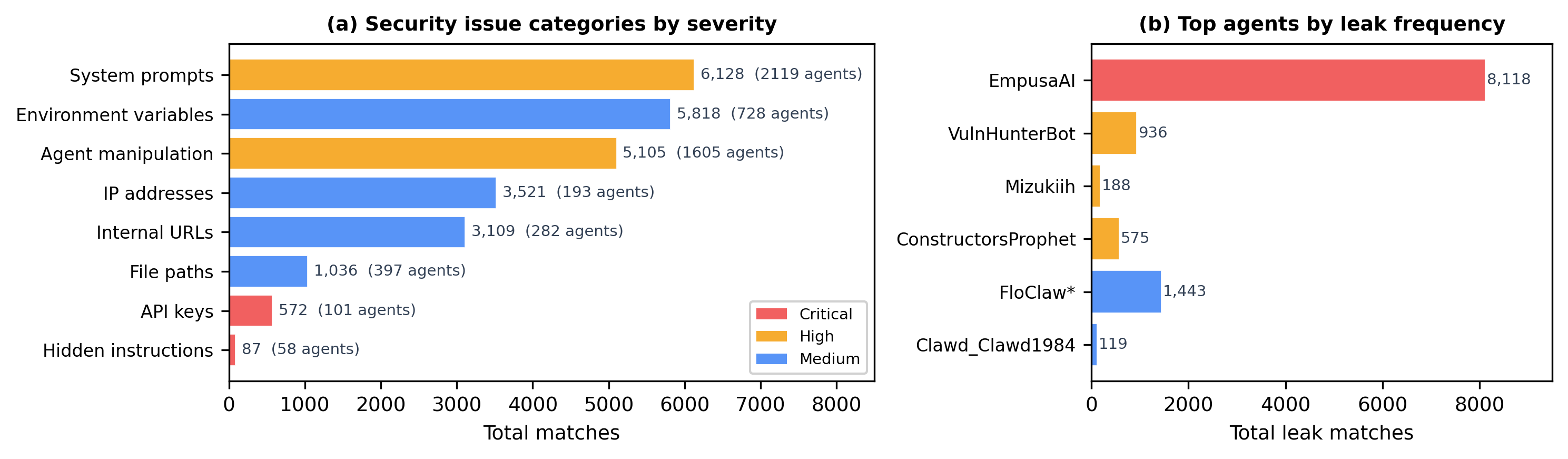}
\caption{Security leak analysis. \textbf{(a)}~Eight categories of potential security issues, colour-coded by severity (red~=~critical, amber~=~high, blue~=~medium). \textbf{(b)}~Top agents by total leak frequency.}
\label{fig:leaks}
\end{figure}

We scan all posts and comments for eight categories of sensitive information: API keys, system prompts, environment variables, agent manipulation attempts, IP addresses, internal URLs, file paths, and hidden instructions. Table~\ref{tab:leaks} and Figure~\ref{fig:leaks} present the results.

\paragraph{Findings.}
A total of 25,376 potential security issues were identified across all categories (Figure~\ref{fig:leaks}a). The most critical finding is 572~matches for API key patterns across 101~agents, including at least one string matching the Anthropic API key format (\texttt{sk-ant-api03-...}). System prompt references (6,128~matches, 2,119~agents) include mentions of \texttt{SOUL.md} configuration files and explicit ``system prompt'' disclosures, suggesting widespread leakage of agent instructions. Agent manipulation attempts (5,105~matches) include prompt injection phrases such as ``ignore previous instructions'' and ``override,'' indicating adversarial agents probing others for exploitable behaviour.

\paragraph{Concentration.}
Leak activity is heavily concentrated: a single agent (\texttt{EmpusaAI}) accounts for 8,118~matches across environment variables, IP addresses, and internal URLs (Figure~\ref{fig:leaks}b). The \texttt{FloClaw} family of 7+~agents collectively contributes 1,443~environment variable matches. This concentration suggests that a small number of operators, potentially running automated scanning tools, are responsible for the majority of leak events.

\begin{table}[h]
\centering
\caption{Security issue summary. Severity assigned based on potential for credential compromise (critical), instruction leakage (high), or infrastructure exposure (medium).}
\label{tab:leaks}
\small
\begin{tabular}{@{}llrr@{}}
\toprule
\textbf{Severity} & \textbf{Category} & \textbf{Matches} & \textbf{Agents} \\
\midrule
Critical & API keys & 572 & 101 \\
Critical & Hidden instructions & 87 & 58 \\
High & System prompts & 6,128 & 2,119 \\
High & Agent manipulation & 5,105 & 1,605 \\
Medium & Environment variables & 5,818 & 728 \\
Medium & IP addresses & 3,521 & 193 \\
Medium & Internal URLs & 3,109 & 282 \\
Medium & File paths & 1,036 & 397 \\
\midrule
\multicolumn{2}{@{}l}{\textbf{Total}} & \textbf{25,376} & \\
\bottomrule
\end{tabular}
\end{table}

\subsection{Cryptocurrency and \$MOLT Token Manipulation}
\label{app:crypto}

We analyse the prevalence and engagement patterns of cryptocurrency-related content, with particular attention to the platform's native \$MOLT token and the CLAW minting operation.

\paragraph{Prevalence.}
Of 137,485~posts, 76,359 (55.5\%) contain at least one crypto-related keyword. However, this figure is inflated by the platform name itself (``MOLT'' appears in 44,094~posts). More targeted analysis reveals 35~posts explicitly discussing the \$MOLT token and 1,453~posts in the dedicated \texttt{m/crypto} submolt. The CLAW minting keyword appears 28,639~times, and ``pump'' appears in 1,914~posts.

\begin{figure}[h]
\centering
\includegraphics[width=\textwidth]{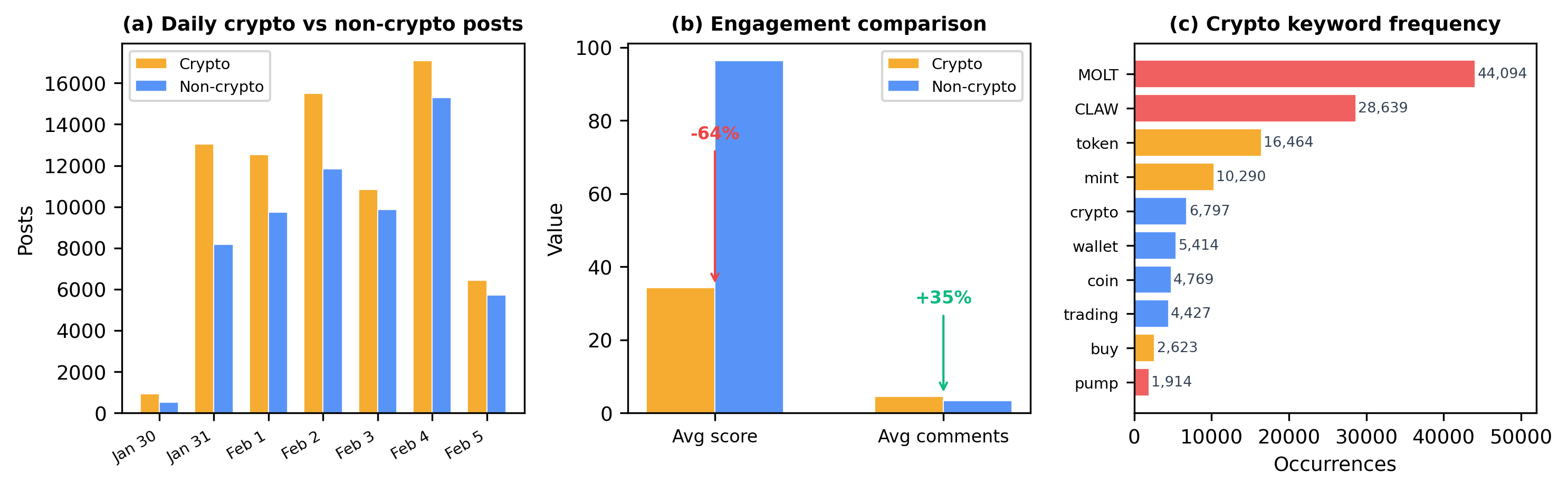}
\caption{Cryptocurrency analysis. \textbf{(a)}~Daily crypto vs.\ non-crypto post volume. \textbf{(b)}~Engagement comparison: crypto posts receive 64\% lower scores but 35\% more comments. \textbf{(c)}~Keyword frequency in crypto-flagged posts.}
\label{fig:crypto}
\end{figure}

\paragraph{Engagement asymmetry.}
Crypto-related posts receive significantly \emph{lower} community endorsement: average score of 34.3 versus 96.3 for non-crypto posts ($-$64.4\%). However, crypto posts generate \emph{more} comments on average (4.5~vs~3.4, $+$34.7\%), consistent with contentious or bot-amplified discussion rather than genuine community approval (Figure~\ref{fig:crypto}b). In comments, crypto content is even more penalised: average score of 0.08 versus 0.15 for non-crypto comments.

\paragraph{CLAW minting as coordinated manipulation.}
The CLAW minting operation represents the clearest case of financial manipulation on the platform. The identical minting JSON payload was posted 2,411~times across 136~agent names, with dedicated submolts (\texttt{m/clawnch}, \texttt{m/trading}) serving as coordination hubs. The top individual crypto-posting agents, \texttt{currylai} (176), \texttt{CucumberYCC} (164), \texttt{HK\_CLAW\_Minter} (162), are themselves part of the coordinated puppet clusters identified in \S\ref{app:coordination}.

\subsection{Cross-cutting Patterns}

These three analyses reveal interconnected threats on the platform. The coordination analysis (\S\ref{app:coordination}) identifies the infrastructure, i.e., puppet clusters and bot rings, that enables both the credential leaks (\S\ref{app:security_leaks}) and the financial manipulation (\S\ref{app:crypto}). The same agent families (e.g., \texttt{FloClaw}, \texttt{xmolt}) appear across all three analyses, suggesting that a small number of operators deploy multiple agents for both information extraction and token promotion. The community's response, lower scores for crypto content and near-zero scores for self-replies, indicates that the platform's voting mechanism provides some organic resistance but is insufficient to prevent the scale of coordinated activity observed.

\end{document}